\documentclass[11pt,a4paper]{iopart}
\usepackage{graphicx,color}
\usepackage{dcolumn}
\usepackage{epsfig}
\usepackage{setstack}

\usepackage{iopams}
\usepackage{euscript}
\usepackage{amssymb}
\usepackage{amsthm}
\usepackage{newlfont}
\usepackage{mathrsfs}
\usepackage{subfigure}
\usepackage{todonotes}
\usepackage{hyperref}

\newcommand{\ie}{\textit{i.e.}}

\def\bea{\begin{eqnarray}}
\def\eea{\end{eqnarray}}
\def\ba{\begin{array}}
\def\ea{\end{array}}
\def\n{\nonumber}
\def\la{\langle}
\def\ra{\rangle}

\def\lsim{\,\lower2truept\hbox{${< \atop\hbox{\raise4truept\hbox{$\sim$}}}$}\,}
\def\gsim{\,\lower2truept\hbox{${> \atop\hbox{\raise4truept\hbox{$\sim$}}}$}\,}

\begin{document}

\title{Active gating: rocking diffusion channels}
\author{Tirthankar Banerjee}
\address{Instituut voor Theoretische Fysica, KU Leuven, Belgium}
\hspace{2.4cm} \href{mailto:tirthankar.banerjee@kuleuven.be}{tirthankar.banerjee@kuleuven.be}
\author{Christian Maes}
\address{Instituut voor Theoretische Fysica, KU Leuven, Belgium}
\hspace{2.4cm} \href{mailto:christian.maes@kuleuven.be}{christian.maes@kuleuven.be}

\begin{abstract}
When the contacts of an open system flip between different reservoirs, the resulting nonequilibrium shows increased dynamical activity.  We investigate such active gating for one-dimensional symmetric (SEP) and asymmetric (ASEP) exclusion models where the left/right boundary rates for entrance and exit of particles are exchanged at random times. Such rocking makes SEP spatially symmetric and on average there is no boundary driving; yet the entropy production increases in the rocking rate.   For ASEP a non-monotone density profile can be obtained with particles clustering at the edges. In the totally asymmetric case, there is a bulk transition to a maximal current phase as the rocking exceeds a finite threshold, depending on the boundary rates. We study the resulting density profiles and current as functions of the rocking rate.
\end{abstract}
\maketitle
\section{Introduction}
\begin{figure}[th]
	\centering
	\includegraphics[width=9 cm]{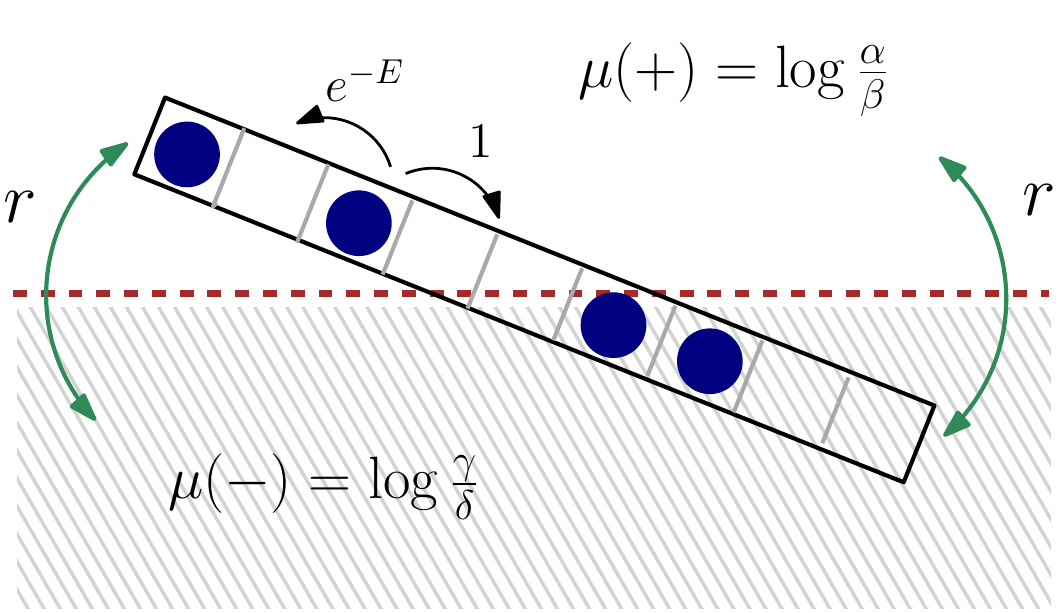}
	\caption{Rocking an exclusion process with external field $E$ in a two-dimensional environment. The switching of contacts happens at rate $r$ with entrance rates $\alpha,\gamma$ and exit rates $\beta,\delta$.  The chemical potentials $\mu(+)$ and $\mu(-)$ correspond to the reservoirs, respectively above and below the dashed line, see Eq.~(\ref{chp}).}
	\label{fig:asep}
\end{figure}

Time-dependence of physical parameters often enters the Hamiltonian or only concerns the bulk dynamics of a system. The present paper deals with the less common situation  where the boundary dynamics is time-dependent, subject in fact to dichotomous noise~\cite{han}.    We will see that the noisy gating, which we implement as ``rocking the system,'' adds spatial symmetry but also dissipation and the extra activity may change the phase diagram of the original process where the boundary condition is fixed.\\
The resulting systems have a (bulk) particle-conserving dynamics with stochastic boundary conditions.  To be specific we consider one-dimensional channels with particle hopping subject to exclusion, as in boundary-driven lattice gases~\cite{Liggett, Spohnbook,zia,ep}.
The boundaries let particles in and out, but the entry and exit rates flip between two values at exponentially distributed times. Thinking of the edges of the system as gates for flows of particles, the models implement {\it active gating}.  Fig.~\ref{fig:asep} shows a possible scenario of rocking the channel; its contacts are randomly flipping between two particle-reservoirs. 
Such modeling appears relevant especially for quasi-one dimensional channels, pores or polymers sitting or moving at the interface between two chemical reservoirs, as is ubiquitous in biological environments.  We also think of pores connecting chemical reservoirs, such as the interior and exterior of a biological cell where however the effective chemical potentials or kinetic parameters are noisy.  The models can also be seen as implementing a version of boundary tumbling, in analogy with active particle processes.  Similar models and ideas have appeared in other contexts, e.g. in \cite{ban,yu} for mathematical aspects.  On the experimental side, the models appear related to the bio- and chemical physics in e.g. \cite{fanzhang-chem20,acs-nano}.\\ 

The main results of our work are the following : (i) For symmetric simple exclusion processes (SEP), rocking leads to a zero-current nonequilibrium steady state with a flat density profile and an associated mean entropy production rate that increases with the rocking rate. (ii) For totally asymmetric simple exclusion processes (TASEP), rocking modifies the standard phase diagram in terms of the boundary rates. There appears a transition as the rocking rate increases, to a maximal current phase. In the thermodynamic limit, bulk density profiles are either linear or flat. (iii) For the more general case of partially asymmetric simple exclusion processes (ASEP), non-monotone density profiles appear for small rocking and also the current is non-monotone in the rocking rate.  The mean entropy production rate for ASEP reflects the nature of the steady bulk current, and shows nonmonotonicity for moderate bias, while remaining monotonic in the rocking rate when the bias is either large and very small.

The plan of the paper is as follows. We start in the next section by defining the rocking dynamics for exclusion models on an open lattice interval.  The general question is to investigate the influence of rocking on entropy production, on the density profile and its role in modifying the phase diagram in terms of particle current. In Sec.~\ref{mod1} we analyze the active boundary-driven SEP and follow it up with the discussion of the totally asymmetric situation in Sec.~\ref{mod2}. Next in Sec.~\ref{rasep} we briefly present results for the more general partially asymmetric case. We summarize and conclude in Sec.~\ref{summ}.


\section{Exclusion models with active gating}\label{rtsep}
\begin{figure}[th]
	\centering
	\includegraphics[width=9 cm]{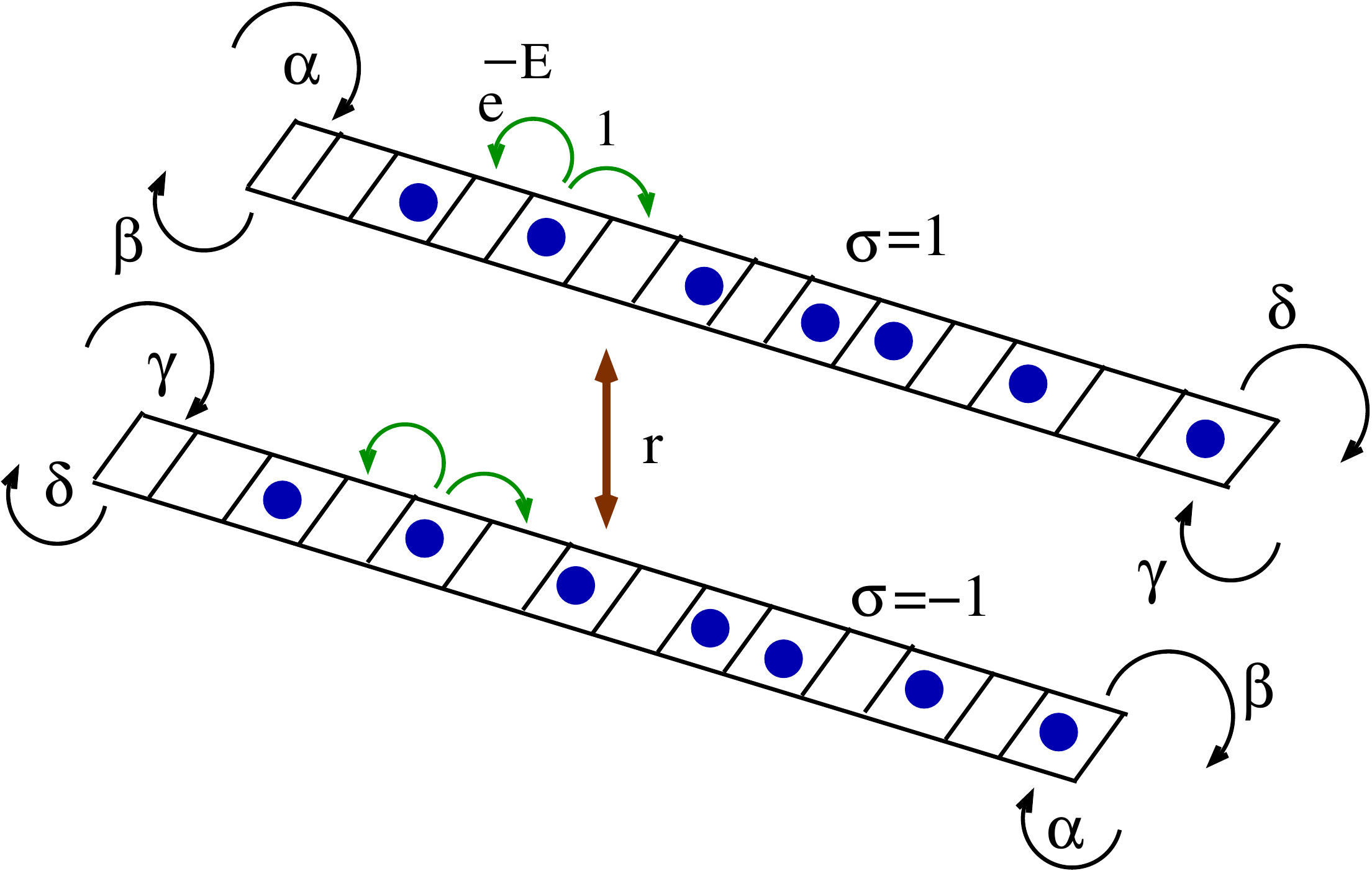}
	\caption{Schematic representation of rocking an exclusion process with external field $E$. Rocking happens at rate $r$ with entrance rates $\alpha,\gamma$ and exit rates $\beta,\delta$.}
	\label{fig:asep1}
\end{figure}

We consider standard exclusion processes on the lattice interval of length $L$ with occupation variables $n = (n_i; i=1,2,\cdots L)$;  $n_i=0,1$ corresponds to the 
$i^{\textrm{th}}$ site being vacant or occupied, respectively. In the bulk, particles hop
at rate 1 to the right and rate $e^{-E}$ to the left, where $E\geq 0$ is an external field.
At the edges, the entrance and exit  rates depend on a binary variable $\sigma=\pm 1$.  We define
\bea
\lambda_{\textrm{in}}(\sigma) := \frac{\alpha+ \gamma}{2} +\sigma\,\frac{\alpha- \gamma}{2} \;= 
\left \{
\begin{array} {cc}
\alpha & \textrm{when} \;\;\sigma =1\\
\gamma & \textrm{when}\;\; \sigma =-1
\end{array}
\right.
\label{fr1}
\eea
and
\bea
\lambda_{\textrm{out}}(\sigma) := \frac{\beta+ \delta}{2} +\sigma\,\frac{\beta- \delta}{2} \;= \left\{
\begin{array} {cc}
\beta &\textrm{when} \;\; \sigma =1\\
	\delta & {\textrm{when}}\;\;\sigma =-1
\end{array}
\right.
\label{fr2}
\eea
for entrance frequencies $\alpha$ and $\gamma$ and exit frequencies $\beta$ and $\delta$, non-negative parameters.
The variable $\sigma$ flips between $\pm 1$ at rocking rate $r>0$; at exponential times the left and the right exchange their entrance/exit rates --- see Fig.~\ref{fig:asep1}.

 Taking into account the exclusion  and the external field $E$ we get for entrance (in) and exit (out) rates,
for the left edge:
\bea
\lambda_{\textrm{in}}^{\textrm{left}}(n_1,\sigma) &=& \lambda_{\textrm{in}}(\sigma) \,(1-n_1)\cr
\lambda_{\textrm{out}}^{\textrm{left}}(n_1,\sigma) &=& \lambda_{\textrm{out}}(\sigma) \,e^{-E}\,n_1
\eea
and for the right edge:
\bea
\lambda_{\textrm{in}}^{\textrm{right}}(n_L,\sigma) &=& \lambda_{\textrm{in}}(-\sigma)\,e^{-E}\, (1-n_L)\cr
\lambda_{\textrm{out}}^{\textrm{right}}(n_L,\sigma) &=& \lambda_{\textrm{out}}(-\sigma) \,n_L
\eea
The environment thus consists of two particle baths, with random switching of contacts between left and right boundaries, and with bulk external field $E$.\\
  A physically useful parameterization is to write
\begin{equation}\label{physp}
\alpha = a_1\,e^{b_1/2},\qquad \beta=a_1 \,e^{-b_1/2},\qquad \gamma= a_2\,e^{b_2/2},\qquad \delta =a_2\,e^{-b_2/2}
\end{equation}
where the $a_i$ are reactivities and the $b_i$ are chemical potentials (up to a factor of $k_BT=1$ that we set one from now on) of the two particle reservoirs.  {Such writing is in accordance with the }condition of local detailed balance~\cite{Christ-ldbarxiv} for the contact with two chemical reservoirs.  The chemical potential of the reservoir which makes contact with the left edge is 
\bea\label{chp}
\mu_\ell(\sigma) = \log\frac{\lambda_{\textrm{in}}(\sigma)}{\lambda_{\textrm{out}}(\sigma)} = \frac{1}{2} \bigg[\log\frac{\alpha\gamma}{\beta\delta} + \sigma\log\frac{\alpha\delta}{\beta\gamma}\bigg]=\left\{
\begin{array}{ll} 
b_1 & \textrm{for} ~\sigma=1 \\
b_2 & \textrm{for} ~\sigma=-1
\end{array}\right.
\eea
depending on the value of $\sigma$. The chemical potential for the right edge is $\mu_r(\sigma) = \mu_\ell(-\sigma)$; see Fig.~\ref{fig:asep}.  There is therefore a variable  thermodynamic force which equals
\bea
F(\sigma) :=E + \frac 1{L} [\mu_\ell(\sigma) - \mu_r(\sigma)]= E+ \frac 1{L}\sigma\log\frac{\alpha\delta}
{\beta\gamma}
\eea
In the stationary state $\la \sigma \ra=0,$ and the average thermodynamic force $\la F(\sigma) \ra =E,$ while $\la \mu_\ell(\sigma) \ra = \la \mu_r(\sigma) \ra = (b_1+b_2)/2,$ \ie, the  chemical potentials at the left and the right edges are equal under stationary averaging.

The expected instantaneous boundary currents into the environment, at the left and right edges respectively, equal
\begin{eqnarray}
J_\ell(\sigma, n_1) &=& \lambda_{\textrm{out}}(\sigma)e^{-E}\,n_1 -\lambda_{\textrm{in}}(\sigma) (1-n_1)\nonumber\\
J_r(\sigma, n_L) &=&  \lambda_{\textrm{out}}(-\sigma)\,n_L -\lambda_{\textrm{in}}(-\sigma)e^{-E}\, (1-n_L) \label{cre}
\end{eqnarray}
given $\sigma,$  $n_1$ and $n_L$. There is also the instantaneous bulk current $J_i$ which is the expected particle flux towards the right between sites $i$ and $i+1$:
\[
J_i(n) = [n_i(1-n_{i+1}) - e^{-E}\,n_{i+1}(1-n_{i})]
\]
They enter the expected entropy production rate $\dot\Sigma(n,\sigma)$, given $(n,\sigma)$, \cite{ep}, as 
\begin{equation}\label{eps}
\fl	\dot\Sigma(n,\sigma) = -  \mu_\ell(\sigma)J_\ell(\sigma,n_1) - \mu_r(\sigma) J_r(\sigma,n_L) + E\, \left[\sum_{i=1}^{L-1} J_i(n) - J_\ell(\sigma,n_1) + J_r(\sigma,n_L)\right]
\end{equation}
The rocking rate $r$ does not appear  explicitly in these equations above.  The standard, unrocked, exclusion process is recovered by fixing $\sigma\equiv 1$ for all times. The mean entropy production rate in the stationary distribution with expectations $ \langle \cdot \rangle $ depending on all parameters is obtained from \eref{eps} by using $ \langle J_\ell(\sigma,n_1) + J_r(\sigma,n_L)\rangle =0, \langle J_i(n)\rangle = j(E,r)$, 
\begin{equation}
\langle \dot\Sigma(n,\sigma) \rangle
= -\frac 1{2} \log\frac{\alpha\delta}
{\beta\gamma}\,\left< \sigma\,[J_\ell(\sigma,n_1)-J_r(\sigma,n_L)] \right> +  (L+1)\, E\,j(E,r)\label{aepr}
\end{equation}
We will see that the rocking leads to a non-zero entropy production even for $E=0$ when there is left/right symmetry in the steady condition.  Also, for $E\neq 0$, blocking one of the edges, e.g. by putting $a_1=0$ in \eref{physp}, still allows for positive entropy production.\\

The rocking exclusion process $(n(t),\sigma_t)$ is Markovian.  Let us look at the time-evolution of the local density $\la n_i\ra_t$. For $i \ne 1,L,$ the density evolves via the usual ASEP equation in the bulk,
\bea
\fl  \partial_t \langle n_i \rangle = \la e^{-E}(n_{i+1} - n_i - n_in_{i+1} + n_in_{i-1}) + n_{i-1} - n_i - n_in_{i-1} + n_in_{i+1}
 \ra_t \qquad \label{bd}
\eea
At the boundaries the evolution equations take a different form, 
\bea
\partial_t \langle n_1 \rangle_t &=&  - \langle J_\ell(\sigma, n_1)\rangle_t + \langle e^{-E} n_2 (1-n_1) - n_1(1-n_2)\rangle_t \cr
\partial_t \langle n_L\rangle_t &=&  -\langle J_r(\sigma, n_L)\rangle_t + \langle n_{L-1}(1-n_L) - e^{-E} n_L (1-n_{L-1})\rangle_t \label{eq:density1}
\eea
The kinetic equations for $\la \sigma n_i\ra_t$ are obtained similarly. At the boundary sites $i=1,L$,
\bea
\fl \partial_t \la \sigma n_1 \ra_t =  \la \lambda_{\textrm{in}}(\sigma) \sigma (1-n_1) \ra_t - e^{-E}\la \lambda_{\textrm{out}}(\sigma) \sigma n_1 \ra_t \cr
 +\;  e^{-E}\,\la \sigma n_2(1-n_1) \ra_t - \la \sigma n_1(1-n_2) \ra_t- 2 r \la \sigma n_1\ra_t \cr
\fl \partial_t \la \sigma n_L \ra_t  =  e^{-E}\la \lambda_{\textrm{in}}(-\sigma) \sigma (1-n_L) \ra_t - \la \lambda_{\textrm{out}}(-\sigma) \sigma n_L \ra_t  \label{eq:t_n_s} \\ +\; \la \sigma n_{L-1}(1-n_L) \ra_t - e^{-E}\la \sigma n_L(1-n_{L-1}) \ra_t - 2 r \la \sigma n_L\ra_t
\nonumber
\eea
In the bulk, \ie, for $i=2,3, \cdots, L-1$ we have, 
\bea
\fl \partial_t \la \sigma n_i \ra_t =   e^{-E}[\la\sigma n_{i+1}(1-n_i)\ra_t - \la \sigma n_i(1-n_{i-1})\ra_t ] \cr
+  \la \sigma n_{i-1}(1-n_i)\ra_t  - \la \sigma n_i(1-n_{i+1})\ra_t
 - 2r \la \sigma n_{i} \ra_t \label{eq:t_ns_bulk}
\eea
These equations \eref{bd}--\eref{eq:t_ns_bulk} are closed for $E=0$ but not otherwise, {and as we shall see, leads to a full determination of the density for rocking SEP (rSEP).}   We assume that the system starts from an arbitrary density profile and $\sigma=\pm 1$ with equal probability. However, we are primarily interested in the stationary process properties, which are independent of the initial condition. \\

Special cases and limits include:
\begin{enumerate}
	\item
rSEP:  $E=0$. We discuss it in Section \ref{mod1}.  The bulk dynamics is unbiased, and there is left/right reflection symmetry in the stationary particle process for any $r>0$.
\item
rTASEP:  $E=+\infty$, is the subject of Section \ref{mod2}.  The dynamics is totally asymmetric with a particle current from left to right.
\item
fast rocking rate:  $r\uparrow \infty$.   When the reservoirs switch at an infinite pace, the flipping $\sigma_t$ decouples from the particle dynamics $n(t)$ at all finite times $t$.  The process restricted to $n(t)$ becomes formally equivalent to a standard open (TA)SEP with rates $\lambda_{\textrm{in}}^{\textrm{left}} = (1-n_1)\,(\alpha+\gamma)/2,\,$ $\lambda_{\textrm{in}}^{\textrm{right}} =e^{-E}(1-n_L)(\alpha+\gamma)/2$ for the entrance and $\lambda_{\textrm{out}}^{\textrm{left}}= n_1\,e^{-E}(\beta +\delta)/2,\,$ $ \lambda_{\textrm{out}}^{\textrm{right}} = n_L \,(\beta+\delta)/2$ for the exit of particles.  It is as if left and right edges are in contact with the same (nonexisting) reservoir at chemical potential $\log (\alpha+\gamma)(\beta+\delta)^{-1}$.  Observe however that the limiting (physical) entropy production must still be calculated from adding the dissipation into the reservoirs (two chemical and one mechanical, which remain untouched of course) as in \eref{eps}.

\item
slow rocking rate: $r\downarrow 0$.   The stationary process is the equal linear combination of the two standard (TA)SEPs with respective rates $(\alpha,\beta)$ and $(\gamma,\delta)$. {The limit $r \rightarrow 0$  does not produce a Markov process for $n(t)$.}  The stationary densities add however. Note that the process at $r=0$ is not defined, except as the large-persistence limit $r\downarrow 0$.  That limit is time-reversible when $E=0$, but even then, it is not an equilibrium SEP.
\item
no driving: The model is satisfying detailed balance when the two reservoirs have equal chemical potential, $b_1=b_2$ which means that $\alpha\delta=\beta\gamma$, and there is no bulk external bias, $E=0$. That corresponds to the equilibrium scenario; the stationary system is time-reversible. The equilibrium distribution is a product measure with  density $\frac \alpha{\alpha + \beta}=\frac \gamma{\gamma + \delta}$.  In fact all the thermodynamic observables are independent of $r$ in that case. That does not mean however there is no influence of $r$; the dynamics keeps depending on the rocking (except in the special case when $\alpha =\gamma$ and $\beta = \delta$ where the rocking does not change the dynamics) and all kinetic observables are expected to depend on $r$.


\item
equal chemical potential (but non-zero drive) : we can choose it to be zero, $b_1= b_2 = 0$, which corresponds to $\alpha=\beta, \gamma=\delta$.  That will be the choice in Section \ref{mod2} for rTASEP.
\end{enumerate}

\section{Rocking the boundary driven SEP}\label{mod1}
\begin{figure}[th]
 \centering
 \includegraphics[height=7 cm]{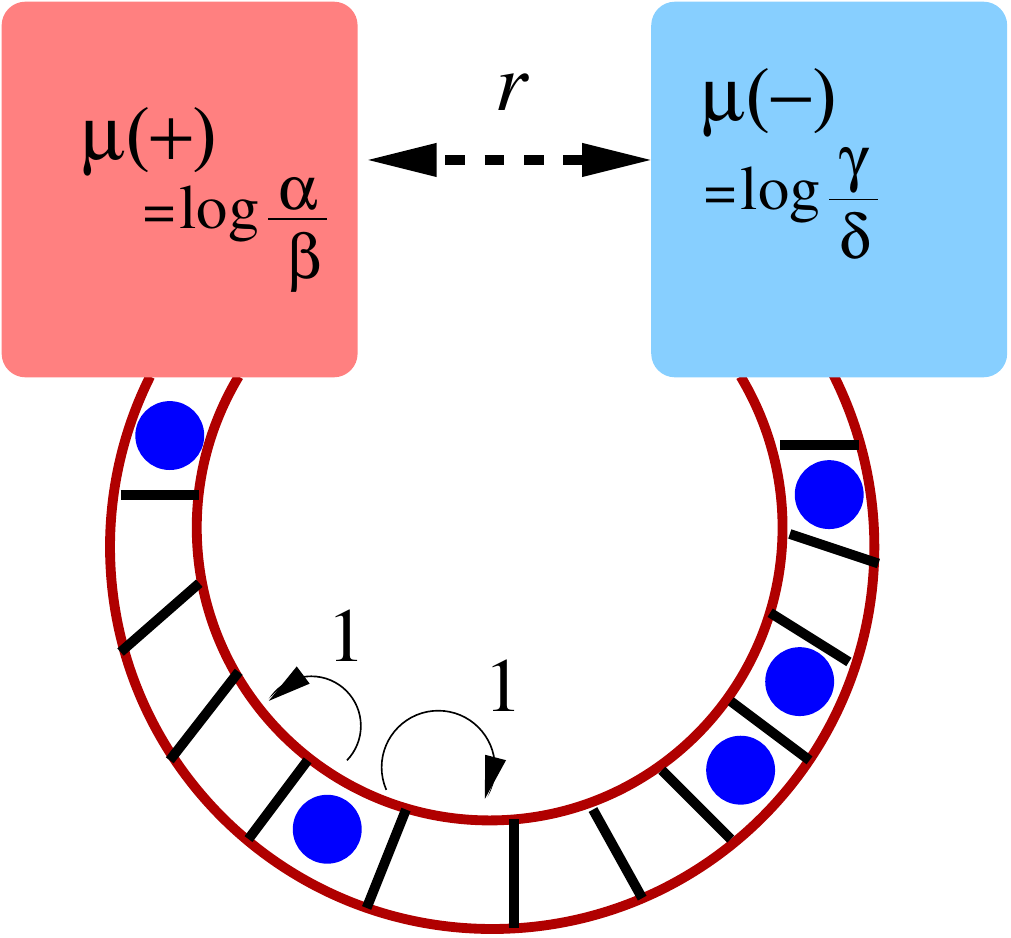}
 \caption{Schematic representation of the rocking SEP. Rocking of gating happens at rate $r$ with entrance rate $\alpha,\gamma$ and exit rates $\beta,\delta$.  }
 \label{fig:sep}
\end{figure}

For the boundary-driven SEP, in the bulk, particles hop symmetrically to neighbouring vacant sites with rate unity.  The boundary rates are determined by \eref{fr1}--\eref{fr2}.   The environment consists of two particle baths, with random switching of contacts with either left or right boundaries as illustrated in Fig.~\ref{fig:sep} (with $E=0$).

\begin{figure}[th]
	\centering
	\includegraphics[width=8.8 cm]{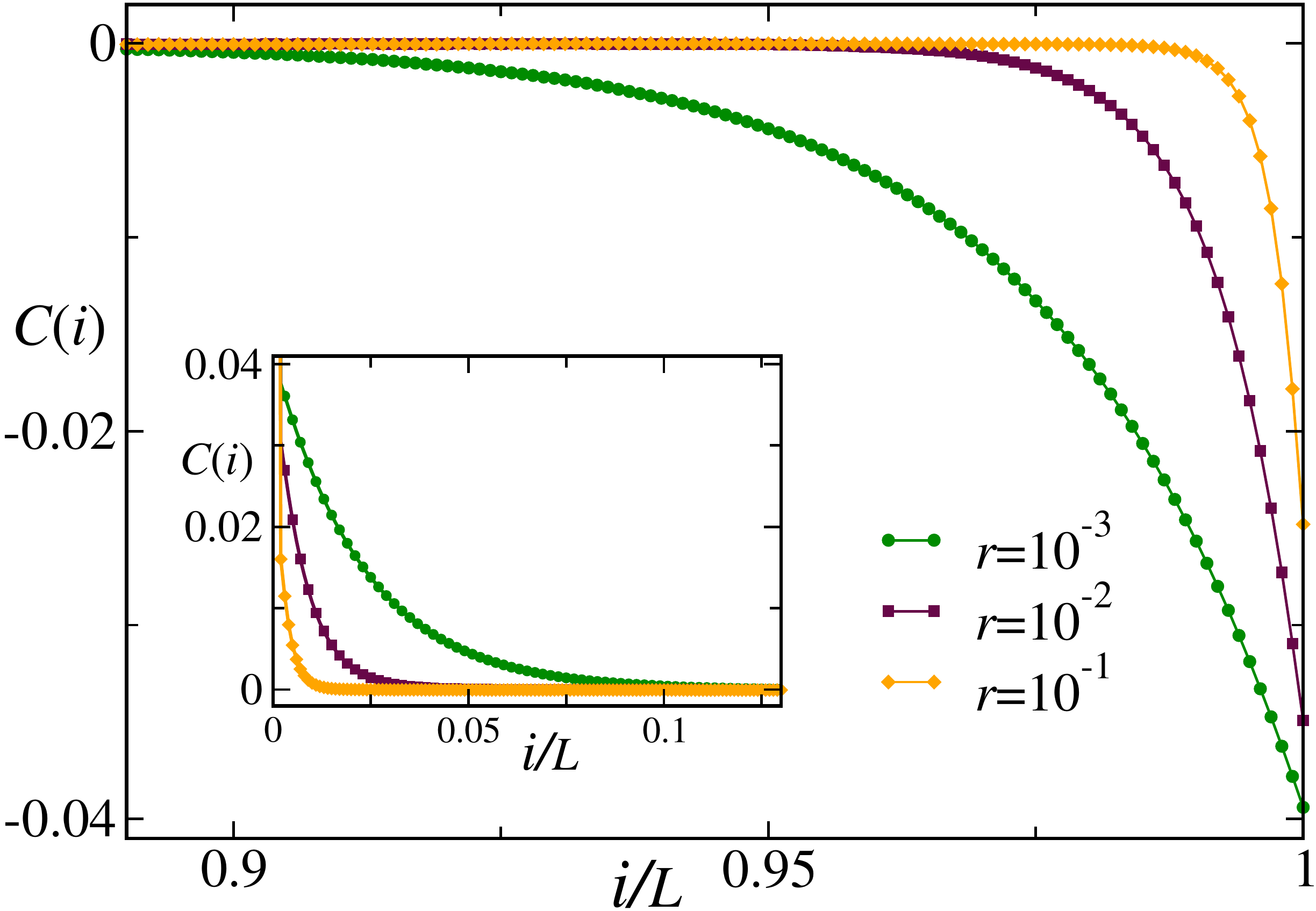}
	\caption{Plot of the spatial correlation $C(i)= \langle n_1 n_i \rangle -\rho^2$ {\em vs} $i/L$ near the right boundary $i=L$ for $L=1000$. The densities at the two boundaries are negatively correlated with the correlations decaying faster for larger $r$. The inset shows the decay of $C(i)$ near $i=1.$  Here $\alpha=0.1$ and $\beta=1=\gamma=\delta.$ Data corresponds to Monte Carlo simulations.
	}
	\label{neg-corr}
\end{figure}

\noindent The random switching of the gates correlates the two edges of the lattice interval. In Fig.\ref{neg-corr} we see how bulk spatial correlations are vanishingly small, while the edge anti-correlation survives of course, in the thermodynamic limit as well. 

\noindent More explicitly, if the left boundary site is empty, a particle enters with rate $\lambda_{\textrm{in}}(\sigma)  = \alpha$ or $=\gamma$ depending on $\sigma = 1$ or $=-1$, respectively (similarly for the right boundary, with entrance rate $\lambda_{\textrm{in}}(-\sigma)$). A particle can leave the system from the left (respectively, right) boundary site with rate $\lambda_{\textrm{out}}(\sigma)$ (respectively, $\lambda_{\textrm{out}}(-\sigma)$). The exit rates switch between $\beta$ and $\delta$.



\subsection{ Density}
The density varies with the edge-parameters and with the rocking rate according to the equations \eref{cre}--\eref{bd} for $E=0$: for $i \ne 1,L,$
\bea
\partial_t  \la n_i \ra_t = \la n_{i-1} \ra_t + \la n_{i+1} \ra_t - 2 \la n_i \ra_t \label{db}
\eea
and at the left and right boundaries,
\bea
\partial_t \langle n_1 \rangle_t &=&  - \langle J_\ell(\sigma, n_1)\rangle_t + \langle n_2 \rangle_t -\langle  n_1\rangle_t \cr
\partial_t \langle n_L\rangle_t &=&  -\langle J_r(\sigma, n_L)\rangle_t + \langle n_{L-1}\rangle_t -\langle n_L\rangle_t \label{eq:density}
\eea
with boundary currents into the environment,
\begin{eqnarray}
J_\ell(\sigma, n_1) &=& \lambda_{\textrm{out}}(\sigma)\,n_1 -\lambda_{\textrm{in}}(\sigma) (1-n_1)\nonumber\\
J_r(\sigma, n_L) &=&  \lambda_{\textrm{out}}(-\sigma)\,n_L -\lambda_{\textrm{in}}(-\sigma) (1-n_L) \label{cres}
\end{eqnarray}
given $\sigma,$  $n_1$ and $n_L$. 
In the stationary state, the lhs of Eqs.~\eref{db}--\eref{eq:density} vanish. 
Since SEP dynamics corresponds to an infinite temperature  and  
the rocking makes the system symmetric under left/right reversal, we expect that the stationary condition has a flat profile $\la n_i \ra =\rho$, independent of $i.$ 

\begin{figure}[ht]
	\centering
	\includegraphics[width=14 cm]{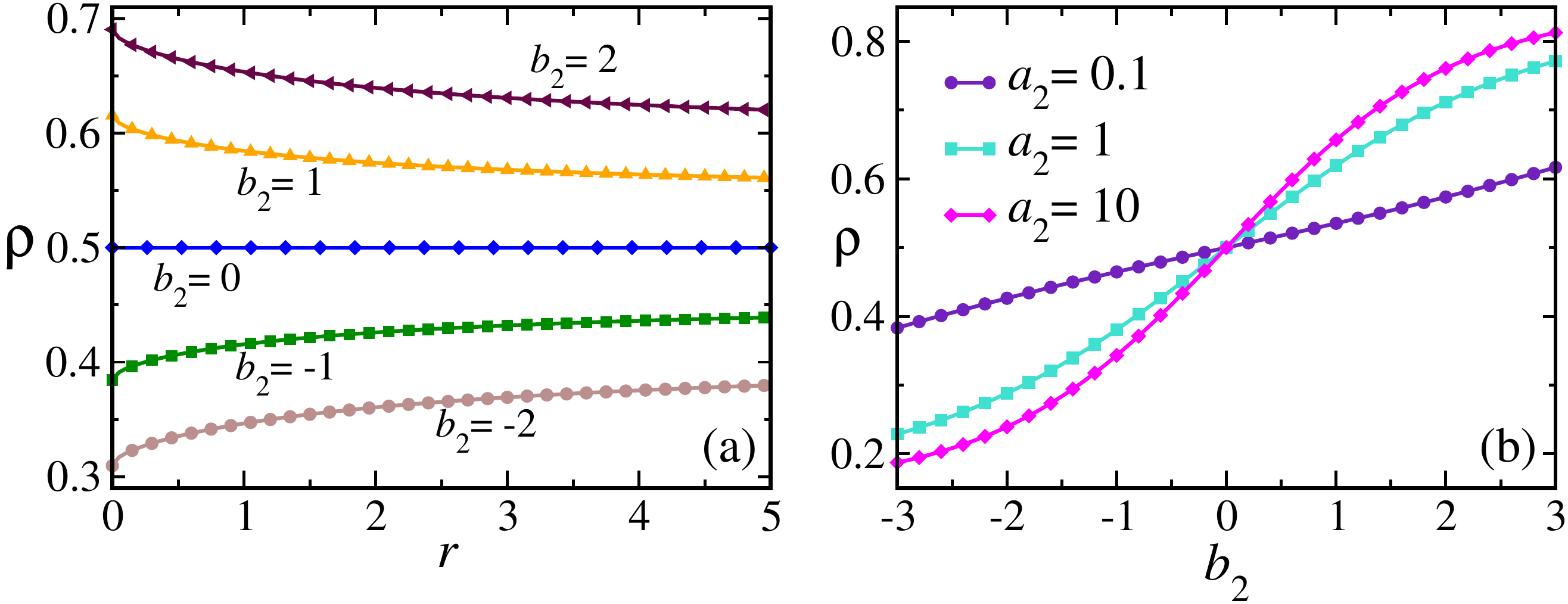}
	\caption{Rocking SEP: (a) Density $\rho$   as a function of $r$ for different values of $b_2.$ Here $b_1=0$ is fixed along with $a_1=5$ and $a_2=1.$ (b) $\rho$ {\it versus} $b_2$ for different values of $a_2$ while $r=1,$  $a_1=1$ and $b_1=0$ are fixed.}
	\label{fig:rho_sep}
\end{figure}

The stationary equations are solved in \ref{recu} for the thermodynamic limit of the density $\rho$.  We find, with $S :=\alpha+\beta+\gamma+\delta, D:= \beta-\delta+\alpha-\gamma $,
and calling $R:= r + \sqrt{r(r+2)}$,
\begin{eqnarray}\label{exr}
\rho &=& \frac{\gamma\beta + \alpha\delta + 2\alpha\gamma + R(\alpha+\gamma)}{2(\alpha+\beta)(\gamma + \delta) + SR}\nonumber\\
 &=& \frac{a_1a_2[e^{(b_2-b_1)/2}+ e^{(b_1-b_2)/2}+2e^{(b_1+b_2)/2}]+ R(a_1e^{b_1/2}+ a_2e^{b_2/2})}{8a_1a_2\,\cosh b_1/2\,\cosh b_2/2 + 2(a_1\cosh b_1/2 + a_2\cosh b_2/2)R}
\end{eqnarray}
which gives the density explicitly in terms of the chemical potentials $b_i$ and the kinetic rates $a_i$. Fig.~\ref{fig:rho_sep}(a) shows the density $\rho$ as a function of $r,$ for different values of the chemical potential $b_2.$  
For $b_1=b_2=b$, we have cancellations giving the equilibrium density
\bea
\rho_{\textrm{eq}}= \frac{e^b}{1+ e^b}. 
\eea
The flat line in Fig.~\ref{fig:rho_sep}(a) corresponds to this equilibrium scenario where the density becomes independent of the kinetic parameters $r,a_1,a_2$.  In general, however, the density depends on these kinetic parameters; see  Figure \ref{fig:rho_sep}(b) where $\rho$ is plotted as a function of $b_2,$ for different values of $a_2.$ A special case is when $b_1 \to -\infty,$ \ie, one of the reservoirs only pumps particle out of the system.  In that case, 
\bea
\lim_{b_1\downarrow -\infty} \rho = \frac{\gamma}{2(\gamma+\delta) + R} = \frac 1{2(1+ e^{-b_2}) + R/a_2} \n
\eea
which remains nonzero for all finite $r,$ and depends explicitly on the reactivity $a_2.$

\subsection{Mean entropy production rate}
\begin{figure}[ht]
	\centering
	\includegraphics[width=14 cm]{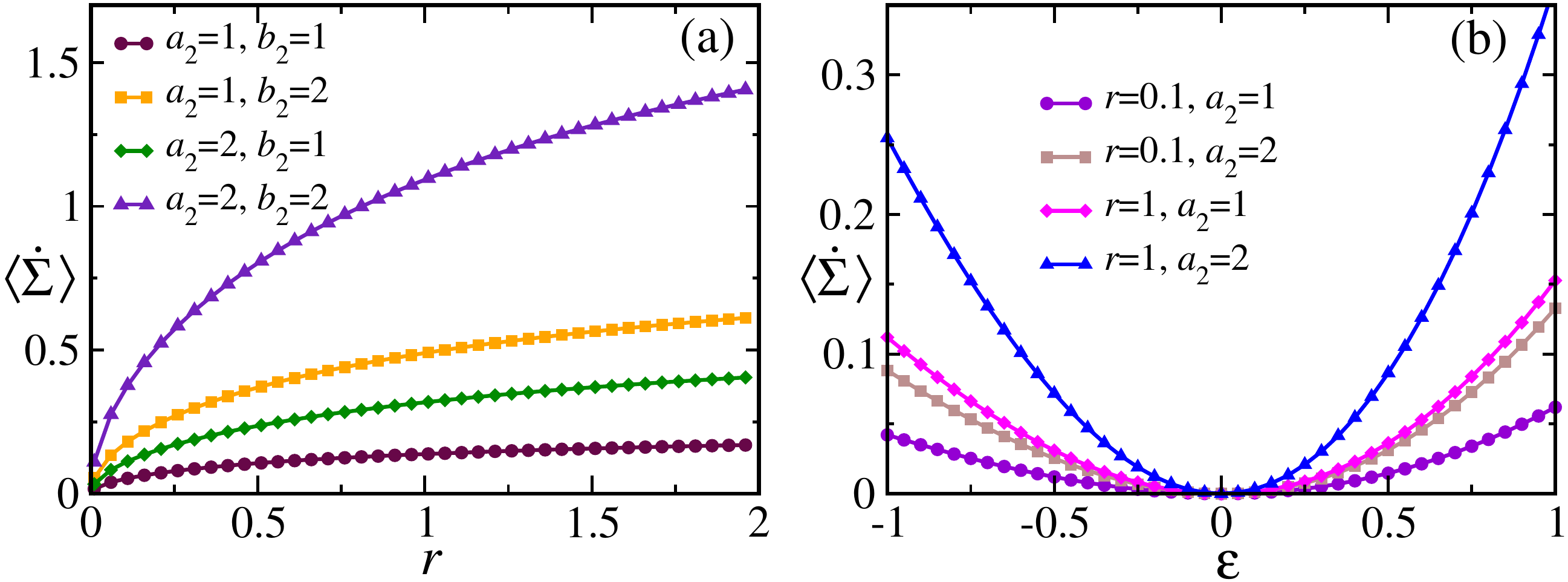}
	\caption{Rocking SEP: (a) Mean entropy production rate $\la \dot \Sigma \ra $ as a function of  rocking rate $r$ for different values of $a_2$ and $b_2.$ Here $a_1=1$ and $b_1=0$ are fixed. (b) $\la \dot \Sigma \ra $  as a function of $\epsilon=b_1-b_2$ for different values of $r$ and $a_2$. $\varepsilon$ is changed by changing $b_2$ while keeping $b_1=1$ fixed.  We have also taken $a_1=1$. $\varepsilon=0$ corresponds to equal chemical potentials at the two gates.} \label{fig:ep_sep}
\end{figure}

It could be surprising that making the boundary driven SEP more spatially symmetric (by the rocking) can increase the entropy production. It does, because of the increased dynamical activity.  The situation is not unlike the one in \cite{onk}, where  a zero-current nonequilibrium state is achieved from SEP from dichotomous stochastic resetting.\\
We can use formula \eref{aepr} to calculate the mean entropy production rate $\langle\dot\Sigma\rangle$. {With $E=0$, we have}
\[
\langle \sigma J_\ell(\sigma,n_1)\rangle = -\langle \sigma J_r(\sigma,n_L)\rangle =
\frac 1{2}(\gamma-\alpha) + \frac{S}{2}\,\langle\sigma\,n_1\rangle + \frac{D \rho}{2}
\]
The result then is
\begin{eqnarray}\label{ep}
\langle \dot\Sigma \rangle &=& \frac 1{2} \log\frac{\beta\gamma}{\alpha\delta}\; \left( \gamma-\alpha + S \,\langle \sigma n_1\rangle + D\,\rho \right)
= \; R\,\langle \sigma n_1\rangle\,\log\frac{\alpha\delta}{\beta\gamma}\cr
&=& \; \frac R D \,(\alpha+\gamma- S\rho) \,\log\frac{\alpha\delta}{\beta\gamma}\nonumber
\end{eqnarray}
Using the exact expression for $\rho$ in \eref{exr}, we have 
\begin{eqnarray}\label{ep1}
\la\dot \Sigma\ra &=& (\alpha \delta - \beta \gamma)\, \log \frac{\alpha \delta}{\beta \gamma} \;\,\frac {1}{S + \frac{2}{R}(\alpha+\beta)(\gamma+\delta)}\\
&=& 2a_1a_2(b_1-b_2)\sinh\frac{b_1-b_2}{2}\, \;\,\frac {1}{S + \frac{8a_1a_2}{R}\,\cosh b_1/2\,\cosh b_2/2}
\end{eqnarray}
Clearly, the entropy production rate is strictly positive whenever $\alpha\delta\neq \beta\gamma$, independent of the system size. It is monotone, increasing with $r$.  The first order in $r\downarrow 0$ in \eref{ep} is purely thermodynamic,
\begin{equation}\label{r0}
\la\dot\Sigma\ra_{r\downarrow 0} = \frac{r (b_1-b_2)}{2\cosh b_1/2\,\cosh b_2/2}\,\sinh\frac{b_1-b_2}{2}
\end{equation}
depending only on the chemical potentials $b_i$.  The maximal mean entropy production rate is reached in the limit
 $r\uparrow \infty$ to give
\begin{equation}\label{rlime}
\la\dot\Sigma\ra_{r\uparrow \infty} = 2\frac{a_1a_2}{S}\,(b_1-b_2)\sinh\frac{b_1-b_2}{2}
\end{equation}
where the kinetics enters via the prefactor $a_1a_2/S$. 
Fig.~\ref{fig:ep_sep}(a) shows the dependence of the mean entropy production rate on the rocking rate $r$  for different values of the chemical potentials and reactivities.

It is also interesting to investigate how the entropy production changes 
 with the chemical potential difference $\varepsilon:= b_1-b_2.$   Figure \ref{fig:ep_sep}(b) shows a plot of $\la \dot \Sigma \ra$ as a function of $\varepsilon$ for different values of $r$ and $a_2.$ For $\varepsilon=0$ the chemical potentials are equal, the system is in equilibrium, and the entropy production vanishes. To quadratic order in  small $\varepsilon$, the mean entropy production rate equals
\begin{equation}
\la\dot\Sigma\ra = \frac{a_1a_2\,\varepsilon^2}{2   \left(a_1+ a_2   + \frac{4a_1a_2 }{R} \cosh \frac{b_1}{2} \right)\cosh \frac {b_1}2}
\end{equation}
and hence keeps kinetic information for all nonzero rocking rates, as  clearly visible from Fig.~\ref{fig:ep_sep}(b).



\section{Rocking TASEP}\label{mod2}


Taking infinite external field $E=+\infty$ in the definitions of Section \ref{rtsep}, we get a rocking TASEP (rTASEP) on an open lattice interval. As we also choose to have constant chemical potential $\mu_\ell=\mu_r=0$ in the environment, the rocking dynamics gets characterized by only two boundary rates  $\alpha=\beta$  and $\delta=\gamma$ and the rocking rate $r>0$. At the left boundary particles only enter; at the right boundary particles can only exit. Apart from using mean field analysis that immediately follows, hereafter, we have also implemented Monte Carlo simulations with random sequential updating to study the system. \\

We start by highlighting a symmetry.  First of all, for every $r>0$ only the external field breaks the left-right symmetry, which can be restored by switching particles and holes and $\sigma\leftrightarrow -\sigma$. That directly leads to
\begin{equation}\label{main-rec-reln}
\langle n_i \rangle = 1- \langle n_{L-i+1} \rangle.
\end{equation}
For example,  $\langle n_1 \rangle + \langle n_L \rangle = 1$ always.  Similarly, $\langle \sigma n_i \rangle = - \langle \sigma\,(1-  n_{L-i+1}) \rangle$ implies
\begin{equation}\label{sn-eq}
\langle \sigma n_1 \rangle = \langle \sigma n_L \rangle
\end{equation}
where all expectations are in the steady state. Note that for all non-zero $r$, the pure low density and high density phases~\cite{evans-asep, krug-prl91} of standard (unrocked) open TASEP vanish. \\
However, the evolution of the bulk density still follows \eref{bd} as for ordinary TASEP ($E \rightarrow \infty$).
At the left and right boundaries we have \eref{eq:density1} and
\begin{eqnarray}\label{tasep-lb}
\partial_t \langle n_1 \rangle_t &=&  \langle J_{\textrm{in}}(\sigma, n_1) \rangle_t - \langle n_1 (1-n_2) \rangle_t \\
\label{tasep-rb}
\partial_t \langle n_L \rangle_t &=& - \langle J_{\textrm{out}}(\sigma, n_L) \rangle_t + \langle n_{L-1} (1-n_L) \rangle_t
\end{eqnarray}
for ({incoming}) and ({outgoing}) boundary currents $
J_{\textrm{in}}$ and $J_{\textrm{out}}$, respectively, with
\begin{eqnarray}
 J_{\textrm{in}}(\sigma, n_1) &=& \frac{1}{2}\left[(\alpha+\delta)-(\alpha+\delta)  n_1 - (\alpha-\delta) \sigma n_1  \right]\label{avg_jl}\\
\label{avg_jr}
J_{\textrm{out}}(\sigma, n_L)  &=& \frac{1}{2}\left[(\alpha+\delta) n_L  - (\alpha-\delta) \sigma n_L  \right]
\end{eqnarray}
In the steady state $j = \la J_{\textrm{in}}(\sigma, n_1) \rangle=\langle J_{\textrm{out}}(\sigma, n_L) \rangle$ is a function of $r$ and a symmetric function of $(\alpha,\delta)$.\\

In the limit $r \uparrow \infty$,
$\langle \sigma n_L \rangle = 0 = \langle \sigma n_1 \rangle$ .  Writing $\lim_{r\uparrow \infty}j = j^{\textrm{sat}}$, \eref{avg_jl} yields
\begin{equation}\label{Jsat}
j^{\textrm{sat}} = \frac{\alpha+\delta}{2}\langle n_L \rangle
\end{equation}
for the saturation current (i.e., for infinitely fast rocking) through the system.
%
When $j^{\textrm{sat}} = 1/4$, then \eref{Jsat} implies
\begin{equation}\label{nL-mc}
\langle n_L \rangle = \frac{1}{2(\alpha+\delta)} = 1-  \langle n_1 \rangle, \qquad \langle n_i(1-n_{i+1})\rangle = 1/4
\end{equation}
Therefore, assuming homogeneous (flat) density in the bulk, under the mean-field approximation $j^{\textrm{sat}} = 1/4$ implies that bulk density to be $\langle n_i\rangle = 1/2$.  The boundary density $\langle n_1\rangle = \langle n_L\rangle = 1/2$ only when $\alpha + \delta = 1$.

\subsection{Phase diagram}

\begin{figure}[th]
 \centering
 \includegraphics[width=7 cm]{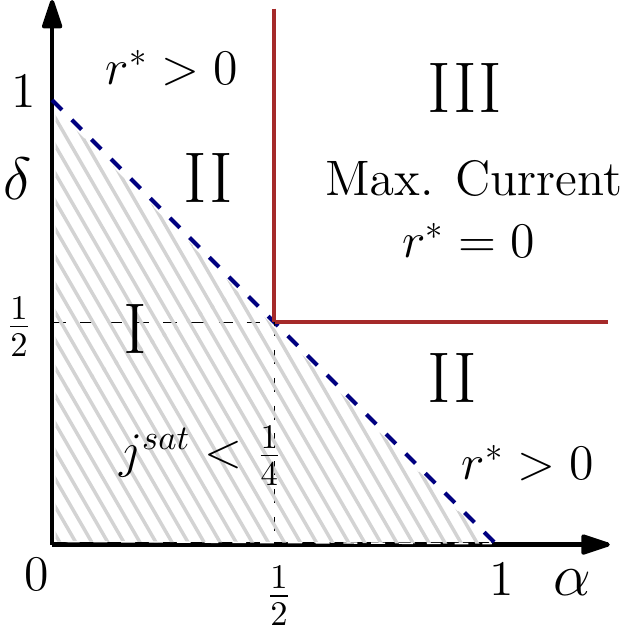}
  \caption{Phase diagram for the rocking TASEP. The phase boundary (dotted line) $\alpha+\delta=1$ separates two regions, characterized by the average steady saturation current $j^{\text{sat}}$ being either less than or equal to $1/4$, respectively in the thermodynamic limit. On the phase boundary itself, the density profile is flat with density $\rho=1/2$ and the current equals $1/4$ in the thermodynamic limit. In region II, the current reaches $1/4$ at finite $r=r^*$. In region III, we have $r^*=0$; see text.}
 \label{pbrinf-fig}
\end{figure}

Figure~\ref{pbrinf-fig} shows how rocking modifies the phase diagram of the usual open TASEP~\cite{evans-asep}.

\noindent It turns out that the line $\alpha +\delta =1$ separates two phases in the thermodynamic limit $L\uparrow \infty$.  The first distinction between those two phases follows from the behavior of the bulk current $j= j(r)$ which increases as $r$ grows :\\
\begin{itemize}
\item Phase A, $\alpha + \delta <1$: strictly monotone in $r$, $j(r)\uparrow j^{\textrm{sat}} < 1/4$. This corresponds to region I in Fig.~\ref{pbrinf-fig}.\\

\item Phase B (regions II and III), $\alpha + \delta > 1$: there exists either  a zero or finite $r^*$ such that $j(r) = 1/4 = j^{\textrm{sat}}$ for all $r> r^*$.   We will see that $r^* =0$ iff both $\alpha,\delta> 1/2$ (region III).  Later a general formula for $r^*$ is given for the case that some rate, say $\alpha < 1/2$ (region II).
\end{itemize}

\begin{figure}[th]
 \centering
 \includegraphics[width=8 cm]{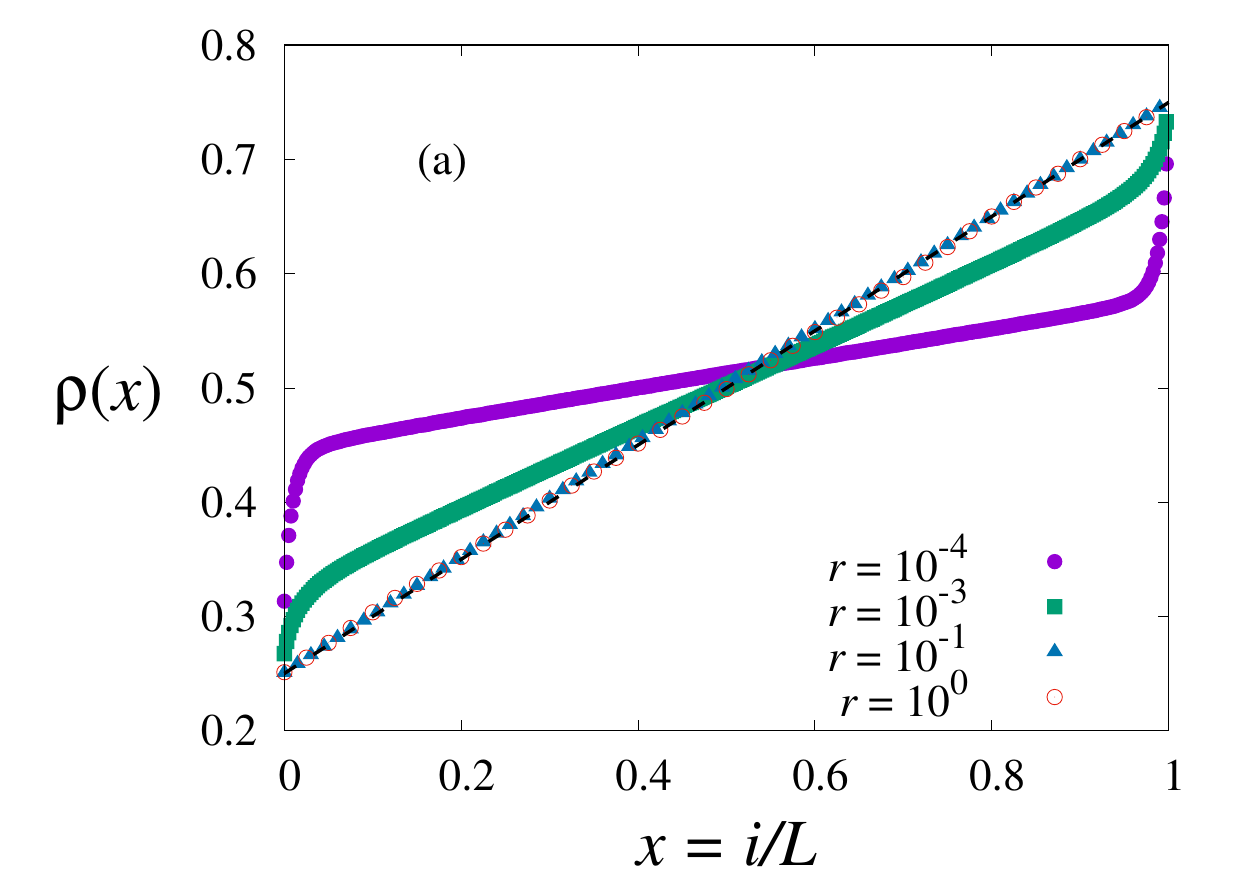}\includegraphics[width=8 cm]{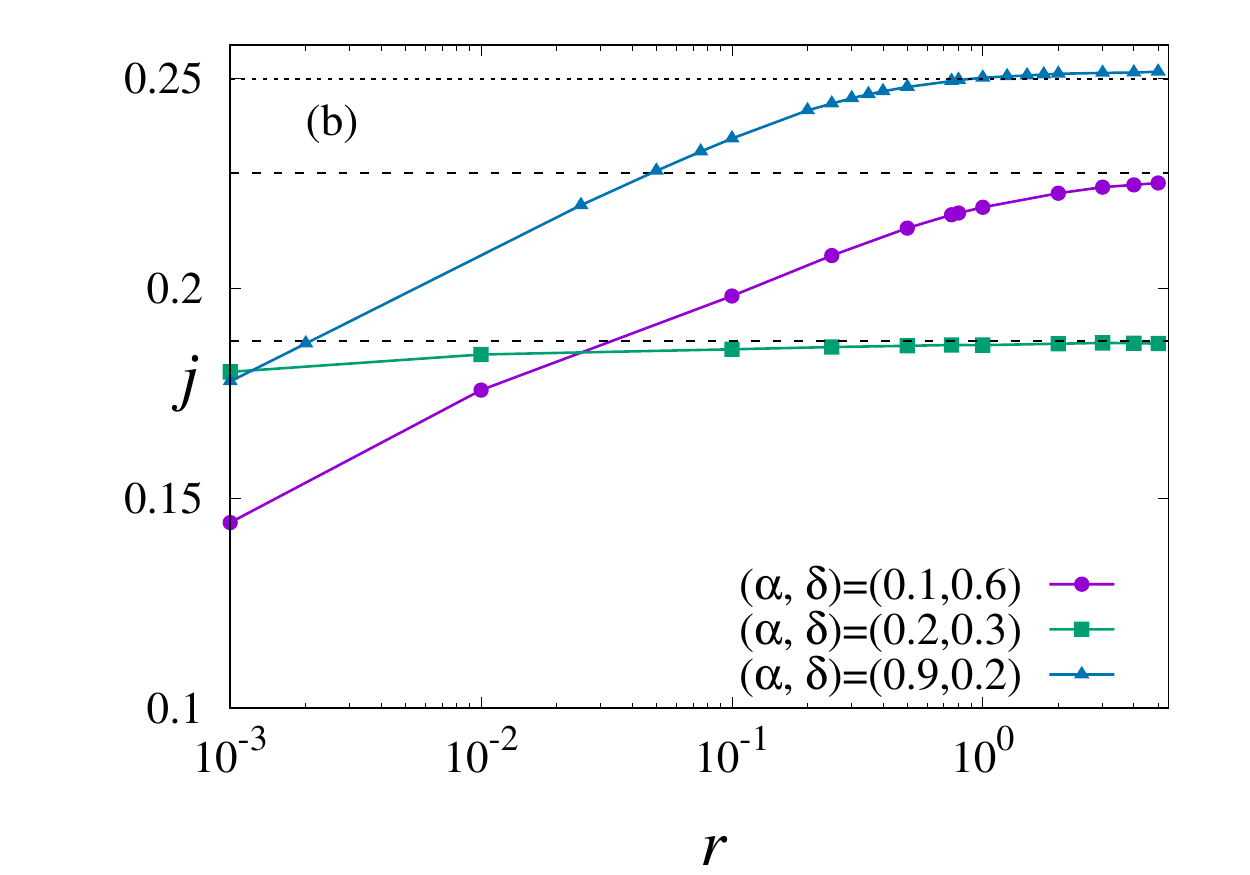}
 \caption{ Rocking TASEP : (a) Plot of average density $\rho(x)$ vs $x=\frac{i}{L}$ for system size $L=400$ for region I : $(\alpha, \delta)=(0.2,0.3)$.  The dashed black line corresponds to the analytical prediction for $r \uparrow \infty$. 
 (b) Plot of average current $j$ vs $r$ in regions I and II. Points represent Monte Carlo results for $L=100$. The dashed black lines refer to the corresponding saturation currents in region I (circles and squares) $j^{\textrm{sat}}=\frac{\alpha+\delta}{2}\left(1-\frac{\alpha+\delta}{2} \right)$. For each case in region I, $j$ approaches $j^{\textrm{sat}}$ as $r$ increases. The dotted line represents the saturation current in region II (triangles) : $j^{\textrm{sat}}= j^{\textrm{max}}=0.25$.}
 \label{den-rI-II} 
\end{figure}

The phase diagram is easiest to understand in the limit $r\uparrow \infty$.
Then, the system feels an {\em effective} rate $f=\frac{\alpha+\delta}{2}$ and connects the left and right edges to densities $f$ and $1-f$, respectively, all the while maintaining the relations (\ref{main-rec-reln}). The average current through the system equals $f(1-f)$. Hence for $\alpha+\delta < 1$, one expects a linear profile for the rocked TASEP that indeed corresponds to moving shocks or delocalized domain walls~\cite{evans-asep}. The linear density profiles connecting left boundary density $f$ to the right boundary density $1-f$ are shown by the dotted lines in Fig.~\ref{den-rI-II}(a). However, when $\alpha+\delta>1$, the system goes to its maximum current phase, complemented with a bulk density equal to $\frac{1}{2}$.\\

\noindent We now dive into the question of how the density and current in each phase region behave as functions of $r$. In what follows, $x$ refers to $i/L$, where $i$ labels the lattice site and $L$ is the system size. We discuss each phase region separately.\\
The phase diagram can be broadly divided into the following three regions :

{Region I} : $\alpha+\delta <1$.


{Region II} : $\alpha+\delta >1$ with $\min\{\alpha,\delta\}<1/2$.

{Region III} : $\alpha+\delta>1$ with both $\alpha, \delta > 1/2$.\\
\par

\begin{itemize}
\item In region I, in the thermodynamic limit, the system shows a linear bulk density profile. The results are shown in Fig.~\ref{den-rI-II}(a).  The density profile always shows a positive slope in the direction of the current. The current is shown in Fig.~\ref{den-rI-II}(b). Assuming $w=\textrm{min}[\alpha ,\delta]$, the average current rises from $w(1-w)$ for $r=0$ to $j^{\textrm{sat}}=\frac{\alpha+\delta}{2} \left(1-\frac{\alpha+\delta}{2} \right)$ as $r$ is increased.
\end{itemize}

\begin{figure}[th]
 \centering
 \includegraphics[width=8 cm]{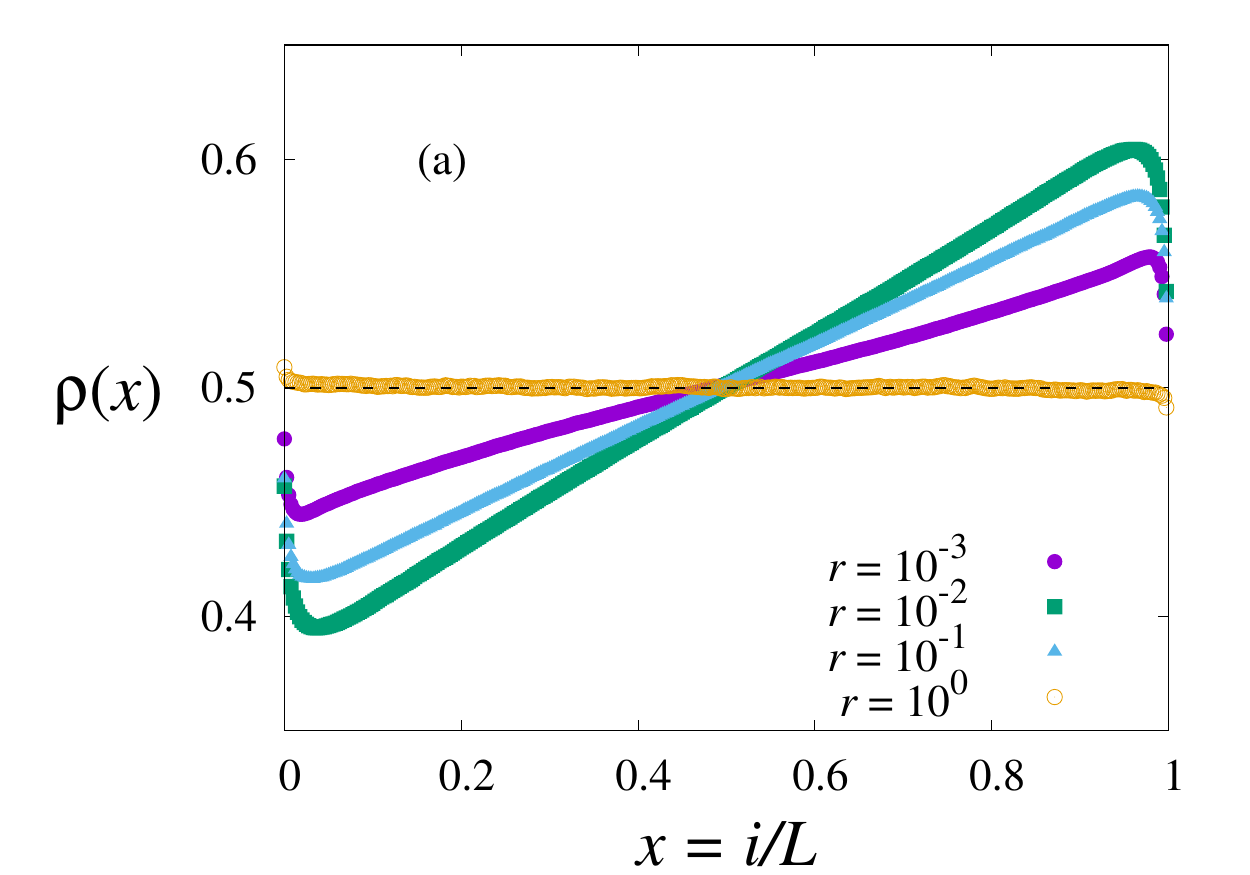}\includegraphics[width=8 cm]{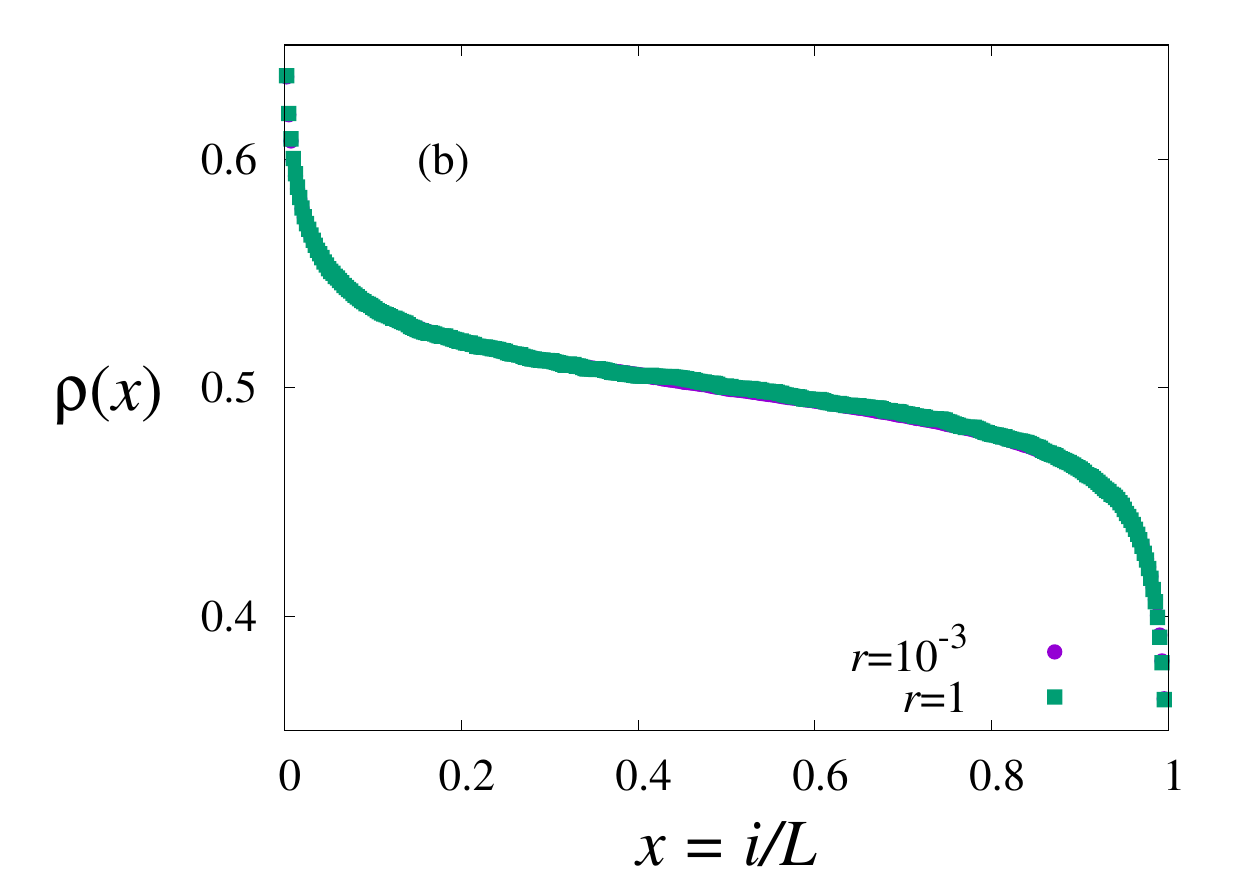}
 \caption{ Rocking TASEP : Plot of average density $\rho(x)$ vs $x=\frac{i}{L}$ for system size $L=400$ (a) in region II: $(\alpha,\delta)=(0.2,0.9)$. For small $r$, the bulk density is linear, while for large enough but {\em finite} $r$, the system goes to its maximal current phase with average current equal to $1/4$. (b) Region III: $(\alpha,\delta)=(0.7,0.8)$. The system mimics the maximal current phase for unrocked TASEP and the density profile remains independent of $r$. The current in the thermodynamic limit is equal to $\frac{1}{4}$ for any $r$ (not shown here). All points represent Monte Carlo results.} 
 \label{den-rIII-IV} 
\end{figure}

\begin{figure}[th]
 \centering
 \includegraphics[width=8 cm]{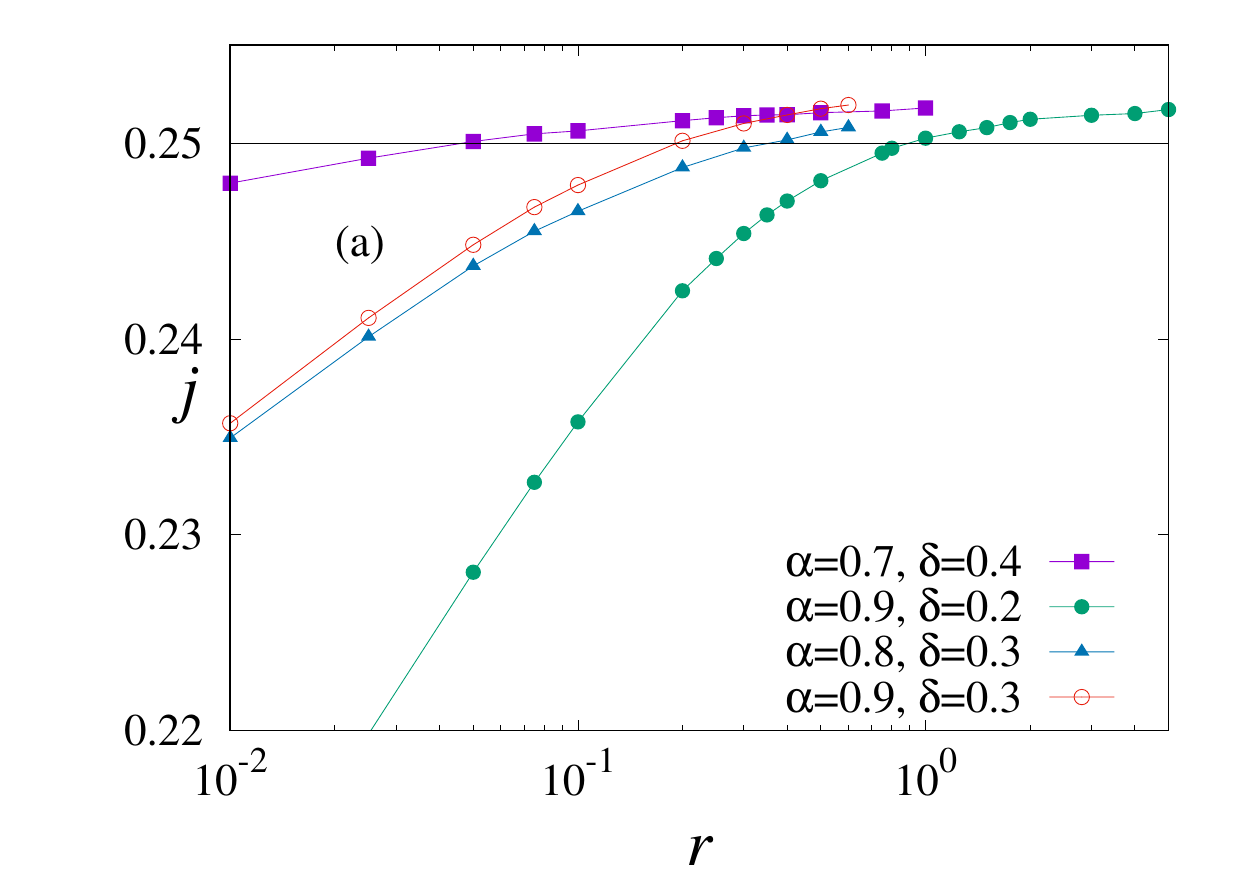}\includegraphics[width=8 cm]{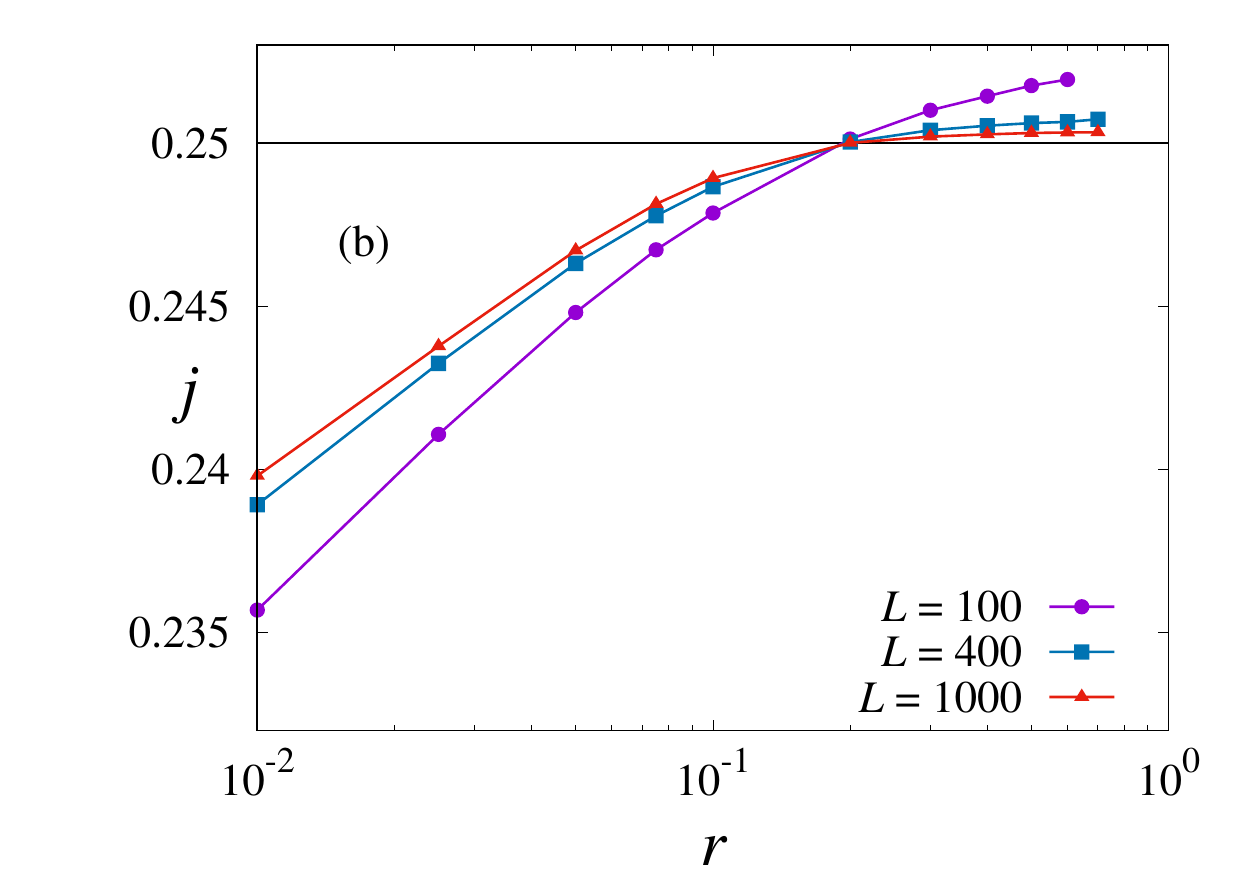}
 \caption{ Rocking TASEP : (a) Plot of $j $ vs $r$ for various values of $\alpha$ and $\delta$ and for $L=100$. For each choice of $\alpha$ and $\delta$, such that $\alpha+\delta >1$ and one of the boundary rates less than $1/2$, we find that there is a finite $r^*$ where the average current through the system reaches $1/4$. (b) Plot of $j$ vs $r$ for  $\alpha=0.3$ and $\delta=0.9$. The value of $r$ at which $j$ reaches $\frac{1}{4}$ is independent of the system size $L$. All points represent Monte Carlo results.}
 \label{jvr1} 
\end{figure}

\begin{itemize}
\item The most interesting aspect of rTASEP resides in region II.
Here the bulk density profile goes from being linear for small $r$ to a constant bulk value $1/2$; see Fig.~\ref{den-rIII-IV}(a). The average steady current, on the other hand rises monotonically with $r$ to saturate at $\frac{1}{4}$ at a finite value of $r=r^*$. Figure~\ref{den-rI-II}(b) also shows a plot of $j$ {\em versus} $r$ in region II.
Figure~\ref{jvr1} shows the behavior of the average current in region II for various choices of $\alpha$ and $\delta$. We discuss region II in the next subsection.
\end{itemize}

\begin{itemize}
\item In region III, the system always shows a flat density profile in the bulk and the current $j(r)=1/4$ always, in the thermodynamic limit for any $r>0$.  Here the system basically mimics the so called maximal current phase of standard open boundary TASEP. An intuitive understanding of this region is as follows: When both boundary rates are larger than $0.5$, rocking essentially gives rise to two new boundary rates which are still larger than $0.5$. Thus, as expected, the rocked system in this limit remains in the maximal current phase ($j=0.25$), independent of $r$; see Fig.~\ref{den-rIII-IV}(b).
\end{itemize}

\subsection{The modified maximal current phase}

In ordinary TASEP, the system reaches its maximal current phase, with $j =\frac{1}{4}$, {\em only} if both the boundary rates are greater than $\frac{1}{2}$. However, when rocking is introduced, the system can show maximum current, even when one of the boundary rates remains lower than $\frac{1}{2}$.  Interestingly, that happens with a transition at a finite rocking value $r=r^*$.\\
We extract now the dependence of $r^*$ on the boundary rates $\alpha$ and $\delta$ in a mean field analysis.\\

\begin{figure}[th]
 \centering
 \includegraphics[width=8.8 cm]{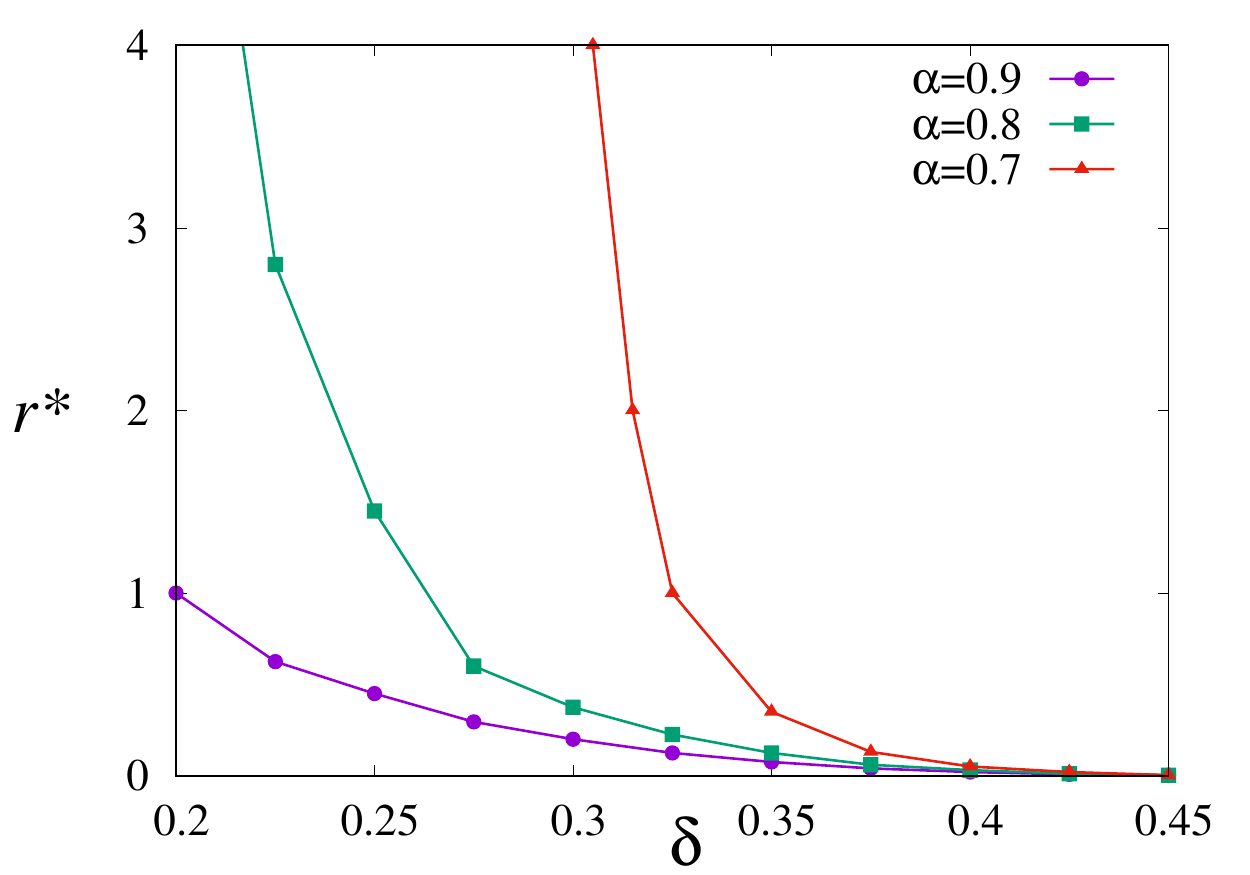}
 \caption{Rocking TASEP : Plot of $r^*$ vs $\delta$  at fixed $\alpha > \frac{1}{2}$ and for $L=100$. $r^*$ increases rapidly as $\alpha+\delta \rightarrow 1$ and for the same value of $\alpha+\delta$, $r^*$ increases with $\alpha-\delta$ as well, as expected from (\ref{rstar-j}). Also, $r^*$ goes towards zero continuously as $\delta \rightarrow \frac{1}{2}$. Points represent Monte Carlo results.}
 \label{rstvbeta} 
\end{figure}

We start with the evolution equation of $\langle\sigma n_i\rangle$. For the boundary sites $i=1, L$, 
\begin{equation}\label{sn1}
\partial_t \langle \sigma n_1 \rangle_t = \langle \lambda(\sigma) \sigma(1-n_1) \rangle_t - \langle n_1 \sigma (1-n_2) \rangle_t - 2r \langle \sigma n_1 \rangle_t
\end{equation}
and
\begin{equation}\label{snL}
\partial_t \langle \sigma n_L \rangle_t = -\langle \lambda(-\sigma) \sigma n_L \rangle_t + \langle n_{L-1} \sigma(1- n_L)\rangle_t - 2r \langle \sigma n_L \rangle_t \,.
\end{equation}
For the bulk, the evolution equations for $\langle \sigma n_i \rangle$ are
\begin{equation}\label{sni}
\partial_t \langle \sigma n_i \rangle_t = \langle  n_{i-1} \sigma (1-n_{i}) \rangle_t - \langle \sigma n_{i} (1-n_{i+1}) \rangle_t - 2r\langle \sigma n_i \rangle_t
\end{equation}
From (\ref{sn1}) in the steady state,
\begin{equation}\label{snL-n1}
(\alpha-\delta) \langle n_L \rangle = [ \alpha+\delta+4r+2\langle n_{L-1} \rangle] \langle \sigma n_1 \rangle ,
\end{equation}
which implies
\begin{equation}\label{sn1-nL}
\langle \sigma n_1 \rangle = \frac{(\alpha-\delta) \langle n_L \rangle}{\alpha+\delta+4r+2 \langle n_{L-1} \rangle},
\end{equation}
where we have also used $\langle \sigma n_1 \rangle = \langle \sigma n_L \rangle$.  Using either (\ref{avg_jl}) or (\ref{avg_jr}), one can write an equation for the current in terms of the rates $\alpha, \delta, r$ and densities $\langle n_L \rangle$ and $ \langle n_{L-1} \rangle$,
\begin{equation}\label{jwithr}
j = \frac{2\alpha\delta+(\alpha+\delta)(2r+\langle n_{L-1} \rangle)}{\alpha+\delta+4r+2\langle n_{L-1} \rangle}\langle n_L \rangle .
\end{equation}
As one would expect, the above relation is symmetric in $\alpha$ and $\delta$.
We are interested in region II of Fig.~\ref{pbrinf-fig}, where $\alpha+\delta >1$ but one of the boundary rates is less than $1/2$. From Fig.~\ref{jvr1} we  see that there exists a finite $r^*$ at which the current $j$ saturates: for small $r$, $j(r)< 1/4$ and it approaches $1/4$ as $r\uparrow r^*$. An expression for $r^*$ in terms of the boundary rates $\alpha$ and $\delta$ ($\alpha+\delta >1$, assuming one of these rates to be less than $0.5$) is obtained by setting $j=1/4$ in (\ref{avg_jl}):
\begin{equation}\label{rstar-j1}
0.5 = (\alpha+\delta)\langle n_L \rangle - \frac{(\alpha-\delta)^2\langle n_L \rangle}{\alpha+\delta+4r+2\langle n_{L-1} \rangle},
\end{equation}
implying
\begin{equation}\label{rstar-j}
r^{*} = \frac{(\alpha-\delta)^2\langle n_L \rangle}{4(\alpha+\delta)\langle n_L \rangle -2} - \frac{\alpha+\delta+2\langle n_{L-1} \rangle}{4}
\end{equation}
Observe that $r^*$ depends not only on $\alpha+\delta$ but also on $(\alpha-\delta)^2$. When the boundary rates are close to each other, one has to rock the system at a much lower rate to reach the maximal current phase. The smaller boundary rate has a dominant say on the steady current. Consider, for example, two sets of rates ($\alpha=0.9, \delta=0.3$) and $(\alpha=0.8, \delta=0.4$). Both these choices have the same value of $\alpha+\delta$. Clearly from (\ref{rstar-j}), the first set would require a larger value of $r^*$ for the system to reach its maximal current phase. Now consider another example when the sets are ($\alpha=0.9, \delta=0.4$) and ($\alpha=0.8, \delta=0.3$) with both choices corresponding to the same value of $\alpha-\delta$. Here the second set would require a larger value of $r^*$ to reach the maximal current phase. One can verify these predictions from Fig.~\ref{rstvbeta}) that shows a plot of $r^*$ {\em vs} $\delta$ for fixed values of $\alpha$. (Note the rapid increase in $r^*$ as $\alpha+\delta \rightarrow 1$.) What these two examples highlight is the fact that the closer the smaller rate is to $0.5$, the smaller $r^*$ one requires to attain the maximal current phase. It must also be stressed that $r^*$ corresponds to the bulk thermodynamic limit.

Using (\ref{nL-mc}) for $\langle n_L \rangle$ indeed gives $r^* \rightarrow \infty$, thus proving consistency of (\ref{rstar-j}) with the results of the earlier discussions in the text. Unfortunately our mean-field analysis does not give a closed expression for $r^*$ in terms of $\alpha$ and $\delta$ beyond~(\ref{rstar-j}).

\subsection{The large persistence regime $r \rightarrow 0$ for a finite size system}
 \begin{figure}[th]
 	\centering
 	\includegraphics[width=8 cm]{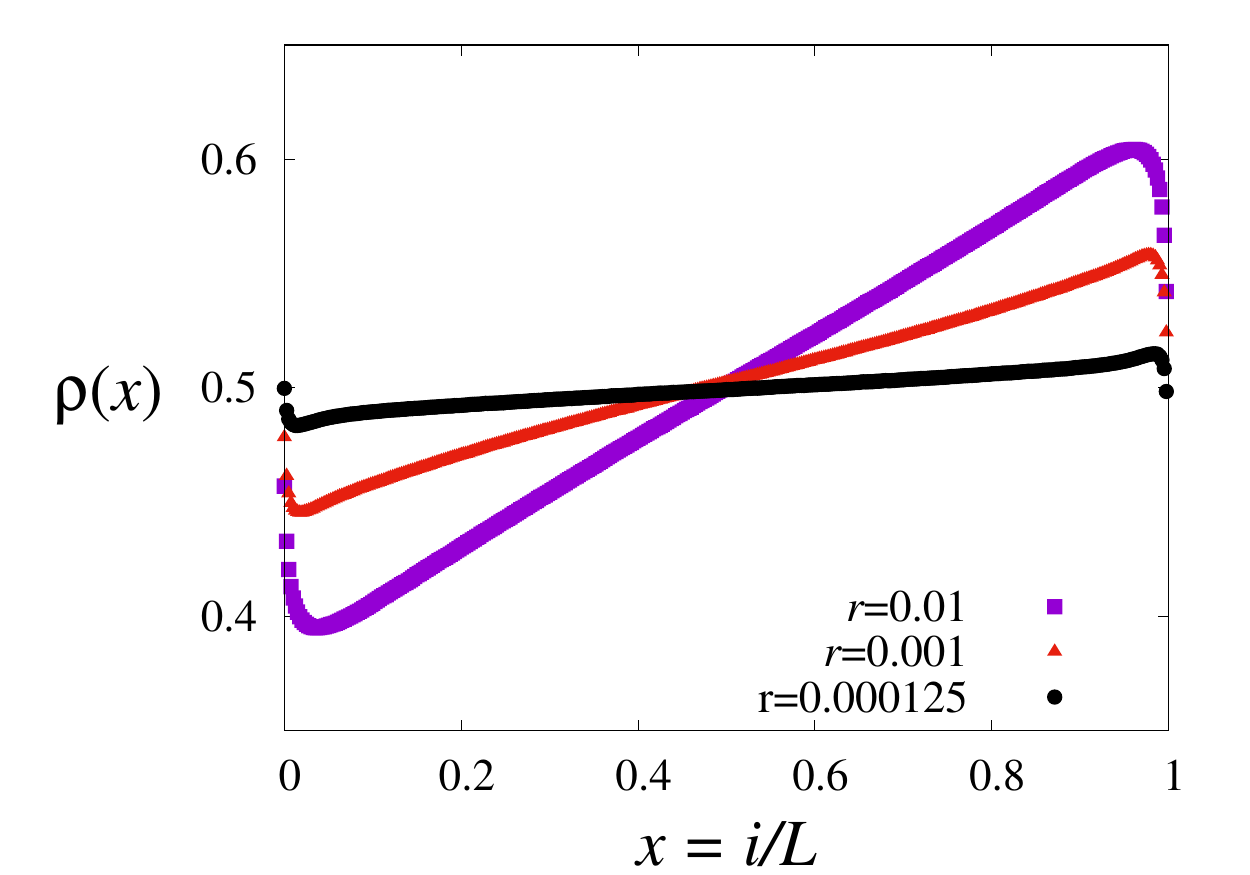}
 	\caption{ Rocking TASEP : $L=400$ and  $(\alpha, \delta)=(0.2,0.9)$ for different values of $r$. Points represent Monte Carlo results. For $r \approx 0$, the bulk density goes to $\frac{1}{2}$, with a negligible slope.} 
 	\label{r0-profile} 
 \end{figure}
 
 At fixed size $L$, when $r\rightarrow 0$, the system mimics the situation where one has two separate TASEP lanes, one with entrance and exit rates $(\alpha, \delta$) and the other with rates ($\delta, \alpha$), respectively. Their densities add up. The density in each lane is determined by $\textrm {min}[\alpha,\delta]$ till one of the rates is less than $0.5$, else the system will only show the maximal current phase. Considering $\alpha < \delta$, we expect the density of the system to be given by $\rho(x) \approx \frac{\alpha+(1-\alpha)}{2} = 0.5$. This relation holds independent of whether $\alpha+\delta$ is greater than or less than $1$. Indeed as $r \approx 0$, the density profile becomes flatter and approaches $0.5$; see Fig.~\ref{r0-profile}.

\noindent The most significant point for large persistence is that the current need not approach $\frac{1}{4}$ even when the bulk density approaches $\frac{1}{2}$. In fact, as $r \rightarrow 0$, the current approaches its value $\alpha(1-\alpha)$; see Fig.~\ref{den-rI-II}(b). This new feature is in sharp contrast to the unrocked TASEP where a bulk density value around $0.5$ would imply a steady current $j \approx \frac{1}{4}$.


 
\section{Rocking ASEP}\label{rasep}
 
 In the presence of a finite bias field $E$,  we are back to the general set up of Section \ref{rtsep}.   The density and current are computed in the limit $r \uparrow \infty$ in \ref{rasep-infty}. \\

\begin{figure}[ht]
	\begin{centering}
			\includegraphics[width=7.5 cm]{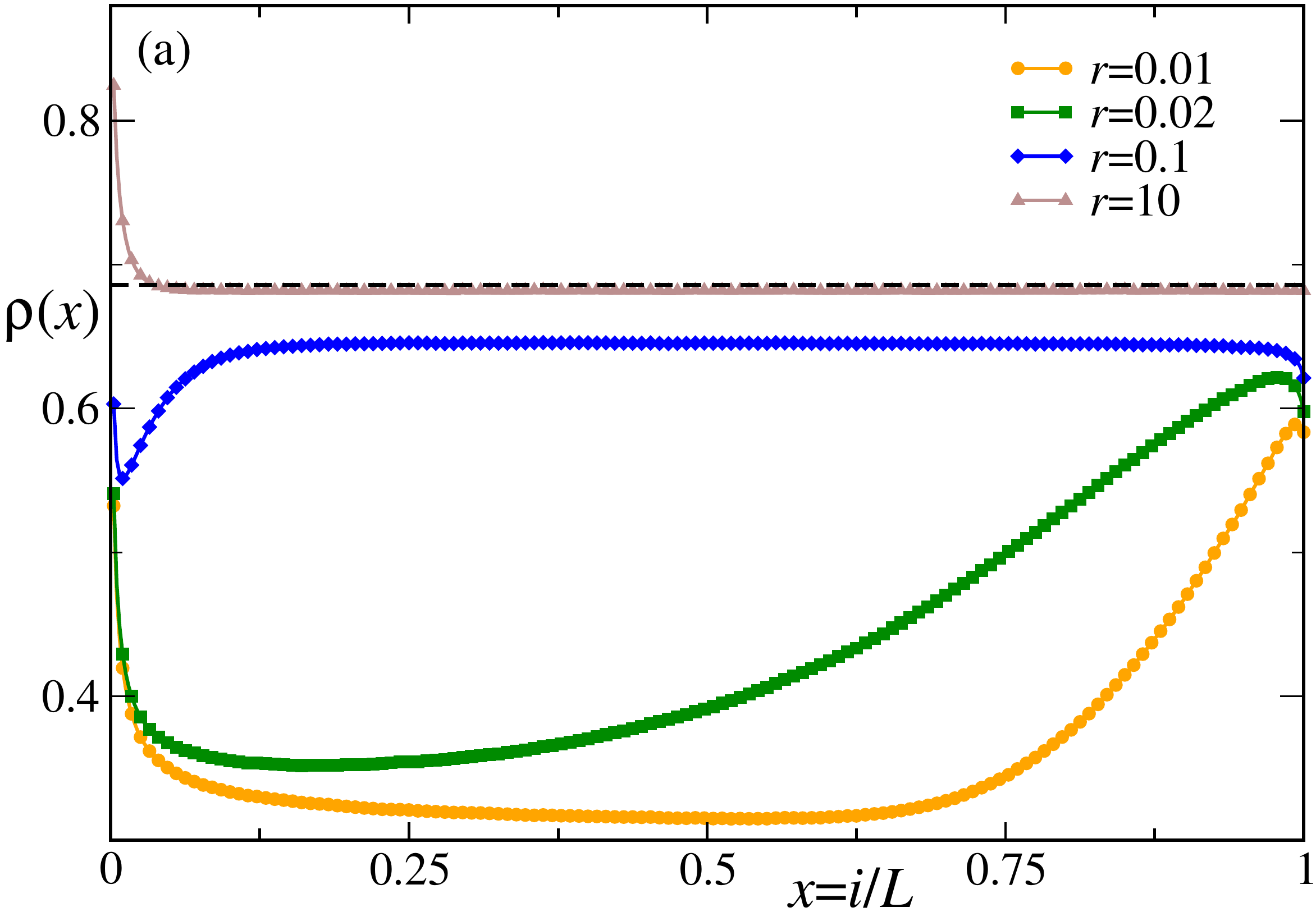}
\hspace*{0.2cm}	\includegraphics[width=7.3cm]{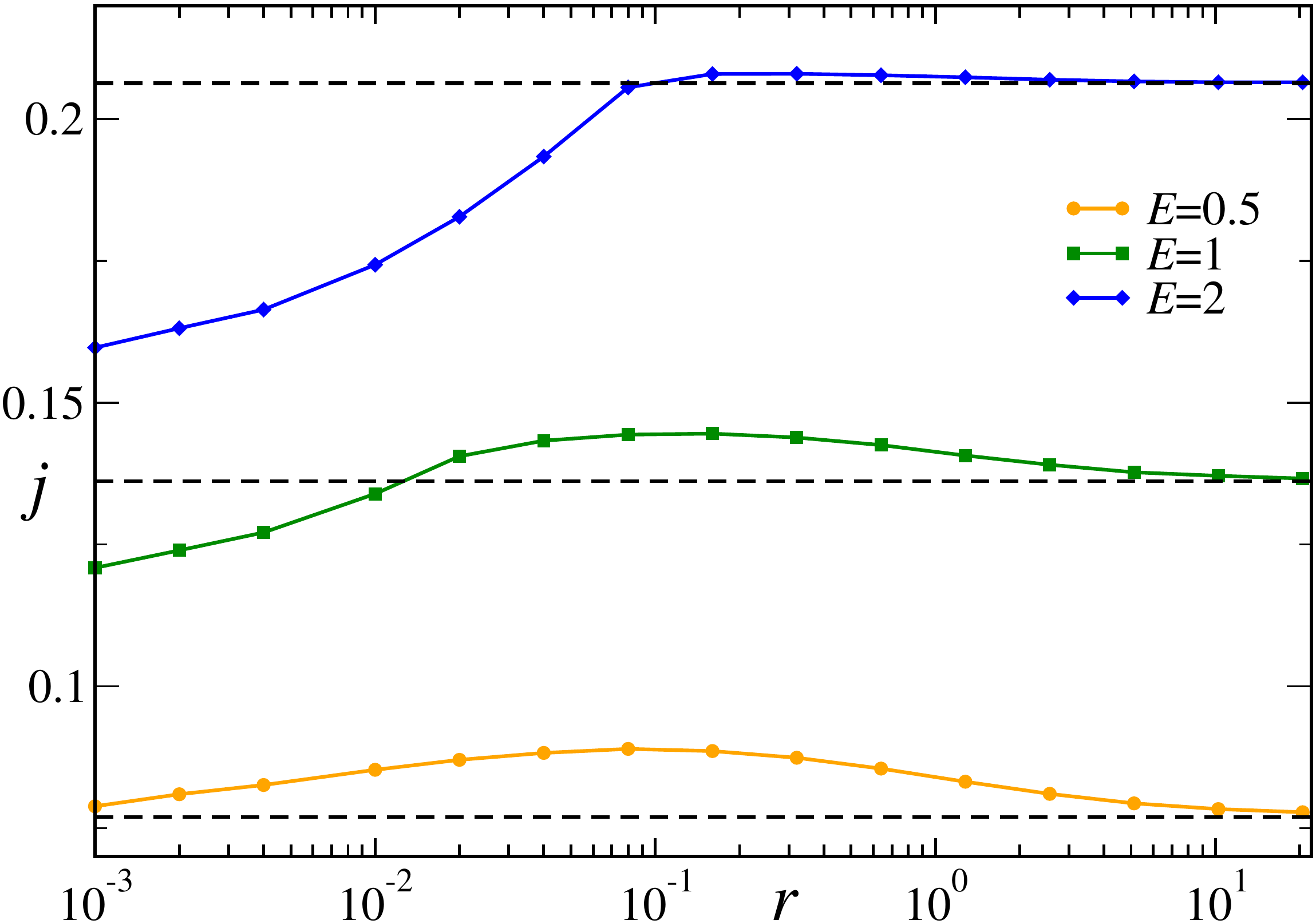}
			
		\caption{rASEP density and current as functions of $r$.   (a) Plot of $\rho(x)$ {\it vs} $x=i/L$ for different values of $r$ and fixed $E=1$. Non-monotonic density profile is observed for $r \downarrow 0$. The dashed line indicates the flat density profile for $r \to \infty$ in the thermodynamic limit. (b) Plot of $j$ vs $r$ for different values of $E.$  The current increases with small $r$, reaches a maximum and then decays to its value for $r \rightarrow \infty$ (indicated by black dashed lines).  The other parameters are $a_1=0.25,a_2=1.25,b_1=-1.5, b_2=2.0$ with system size $L=400.$ Points represent results from Monte Carlo simulations.} 
		\label{asep-HD-den-curr}
	\end{centering}
\end{figure}

\noindent Here we focus first on a particular interesting aspect of rASEP which occurs for small $r$ for choices of the boundary rates that counter-act the external field.  Then, the density profile
shows non-monotonic behavior with particles heaping up at the boundaries.  That is similar to the case of run-and-tumble processes in a trap where the particles tend to cluster at the edges for large enough persistence, \cite{tai}. An example is shown in Fig.~\ref{asep-HD-den-curr} (a).  The non-monotonicity represents the addition of nonlinear densities; for $r\downarrow 0$, the system can either be in a standard ASEP maximal current phase or in a low density phase.
 
\noindent Secondly, the steady current $j$ may be  non-monotone in $r$; see Fig.~\ref{asep-HD-den-curr}(b). {The current increases for small r, reaches a maximum, and then decreases towards its predicted value for $r \uparrow \infty$}(computed  in \ref{rasep-infty}).
The reason is that at small $r$, the system is in a low density phase, with the average density being less than $\frac{1}{2}$, and for $r\uparrow \infty$, the system reaches a high density (average density greater than $\frac{1}{2}$) phase. At low densities the current is small due to very few particles being present in the system, while for high densities, the system progressively gets more jammed. In between these two phases, when the average density is around $0.5$, the current reaches its maximum possible value for the chosen boundary rates.  Note that non-monotone density profiles for TASEP with particle conservation have been observed in~\cite{tirtha-prr} for spatially varying but quenched defects.
 
\noindent  Not shown is the mean entropy production rate as function of $E$ and $r$.  That follows however the same dependencies as the current, from the dominating effect of the $(L+1)j E$ term in Eq.~\ref{aepr}.
%
%
%
%
%
 
 \section{Summary}\label{summ}
We have studied a class of exclusion models (rSEP, rASEP, rTASEP) in one dimension subject to open and {\em active} boundaries. The activity is modeled in terms of a dichotomous noise, where the entry and exit rates at the left boundary flip to their right counterparts at Poisson rate $r$.  Such models can serve as building blocks for further studies on determining efficiency of transport processes~\cite{fanzhang-chem20}, as a function of boundary activity. The tuning of chemical potentials in biological environments can be achieved by monitoring pH gradients~\cite{acs-nano}. \\
	
For rSEP, we find that rocking leads to a zero-current nonequilibrium steady state, characterized by a flat average density throughout and a non-zero entropy production rate that increases monotonically with $r$. For rTASEP, the  phase diagram is modified: the steady current increases with $r$ and then saturates for $r \uparrow \infty$ to either a value less than or equal to $1/4$. Most interestingly, the system can attain its maximal current phase even when one of the boundary rates is less than $1/2$.  The role of rocking is really to increase the smaller rate. When the sum of the rates exceeds $1$, the smaller rate is effectively pushed beyond $1/2$ by the rocking. We also observe that for $r \downarrow 0$, the system always shows a bulk density equal to $1/2$, but where the current need not be equal to $1/4$.  Finally for rASEP, we find that the density profile can be non-monotone for certain boundary rates and for $r \downarrow 0$. The associated current can be a non-monotone function of the rocking rate.\\
The results presented here can be extended in several ways, including more reservoirs or more types of particles. Some boundary parameters can be kept fixed while others vary randomly. How that modifies and selects density profiles and current statistics, remains to be investigated.

\section*{Acknowledgements}

We thank Urna Basu for many discussions and help throughout the course of the work. TB acknowledges support from Internal Funds KU Leuven.
 
\appendix 

\section{Particle density in rSEP}\label{recu}
From Eq.~\eref{cres}, and using the expressions of the rates from Eqs.~\eref{fr1}-\eref{fr2}, we have, in the stationary state,
\bea
\langle J_\ell(\sigma, n_1)\rangle &=& \frac 12 [S \la n_1 \ra + D \la \sigma n_1 \ra -(\alpha + \gamma)] \label{eq:Jlav} \\
\langle J_r(\sigma,n_L)\rangle &=& \frac 12 [S \la n_L \ra - D \la \sigma n_L \ra -(\alpha + \gamma)] \label{eq:Jrav}
\eea
where we denoted $S := \alpha+\beta+\gamma+\delta, D := \beta- \delta+\alpha-\gamma$.  The net current to both the reservoirs must also vanish in the stationary state, \ie, $\langle J_\ell\rangle= \langle J_r\rangle=0$ and we have
\begin{equation}
S\,\rho + D\,\langle \sigma n_1 \rangle= \alpha+ \gamma \label{eq:n_ns}
\end{equation}
for all values of the boundary rates.  From symmetry $\la \sigma n_1 \ra = -\la \sigma n_L \ra$ and both \eref{eq:Jlav} and \eref{eq:Jrav} give rise to the same relation. In the $r\uparrow\infty$-limit we have the decoupling $\la \sigma n_1 \ra = \la \sigma n_L \ra =0$ and hence
\begin{equation}\label{ri}
\langle n_1\rangle_{r\uparrow\infty} = \frac{\alpha + \gamma}{S}
\end{equation}



Equation \eref{eq:n_ns} is not enough to determine the stationary density as it involves the $\la \sigma n_i \ra$ correlation. We need to use the kinetic equations \eref{eq:t_n_s}--\eref{eq:t_ns_bulk} at $E=0$: at the boundary sites $i=1,L$,
\bea
\partial_t \la \sigma n_1 \ra_t  &=& \la \lambda_{\textrm{in}}(\sigma) \sigma (1-n_1) \ra_t + \la \lambda_{\textrm{out}}(\sigma) \sigma n_1 \ra_t +\la \sigma n_2 \ra_t - \la \sigma n_1 \ra- 2 r \la \sigma n_1\ra_t \cr
\partial_t \la \sigma n_L \ra_t  &=& \la \lambda_{\textrm{in}}(-\sigma) \sigma (1-n_L) \ra_t + \la \lambda_{\textrm{out}}(-\sigma) \sigma n_L \ra_t - 2 r \la \sigma n_L\ra_t  \label{t_n_s} 
\eea
In the bulk, \ie, for $i=2,3, \cdots, L-1$, 
\bea
\partial_t \la \sigma n_i \ra_t =  \la \sigma n_{i+1} \ra_t + \la \sigma n_{i-1} \ra_t - 2(r+1) \la \sigma n_{i} \ra_t \label{t_ns_bulk}
\eea
In the stationary state the time-derivatives vanish, and Eqs~\eref{t_n_s} give,
\begin{equation}\label{e2}
\alpha- \gamma - (S+4r)\langle \sigma n_1\rangle -D\,\rho + 2\langle \sigma\,(n_2-n_1)\rangle  =0
\end{equation}
Note that $\la \sigma n_1 \ra = - \la \sigma n_L \ra,$ and the two lines in  Eq.~\eref{t_n_s} reduce to the same relation. In the stationary state Eq.~\eref{t_ns_bulk} gives,
\begin{equation}\label{f2}
\Delta\langle\sigma n_i\rangle =  2r\,\langle\sigma n_i \rangle, \quad i=2,\ldots,L-1
\end{equation}
where $\Delta$ denotes the discrete Laplacian operator. This set of equations is to be solved with the boundary condition $\la \sigma n_1 \ra = - \la \sigma n_L \ra.$ 
To solve the equation \eref{f2},  we write
\begin{equation}\label{eigen}
\Delta f_j := f_{j+1} + f_{j-1} - 2f_j = 2r\,f_j,
\quad j=2,\ldots,L-1
\end{equation}
with boundary condition  $f_L + f_{1} =0$ and $f_j\in [-1,1]$.   Let us take a trial solution $g_j = e^{kj}$  for arbitrary $k.$ 
Then, we must verify
\[
\Delta g_j = e^{kj}(e^{k} + e^{-k} - 2) =  2r g_j
\]
which means that $k$ must be such that
\[
\frac{e^{k} + e^{-k}}{2} - 1 =  r >0
\]
requiring that $\cosh k > 1$ or that $k$ must be real.
Clearly, every linear combination of such $g_j$ still solves the eigenequation.
Per consequence, 
\[
f_j = B\,A^{-1}\,[e^{-k(L+1)/2}\,e^{kj}-  e^{k(L+1)/2}\,e^{-kj}]  = 2B\,A^{-1} \sinh ((j-\frac{L+1}{2})k)
\]
solves the boundary condition and the eigenequation \eref{eigen},  as long as $ A =  2\sinh (\frac{L-1}{2}k)$ and $|B|\leq 1$ to ensure $|f_j|\leq 1$.  In conclusion,
we must have for all $j=1,\ldots, L$,
\[
f_j = B\,\frac{\sinh ((j-\frac{L+1}{2})k)}{\sinh (\frac{L-1}{2}k)},\quad \cosh k= r+1
\]
Obviously,  $f_1 = \langle \sigma n_1\rangle= -B$ fixes $B$:
\begin{equation}\label{sol}
f_j = -\langle \sigma n_1\rangle\,\frac{\sinh ((j-\frac{L+1}{2})k)}{\sinh (\frac{L-1}{2}k)},\quad \cosh k= r+1
\end{equation}
The solution is given by,
\begin{equation}\label{f3}
\langle\sigma n_i \rangle = \langle \sigma n_1\rangle\,\,\frac{\sinh\big[(\frac{L+1}{2}-i)k\big]}{\sinh(\frac{L-1}{2}k)}, \qquad \textrm{where}\; r = \cosh k-1
\end{equation}
For our purposes it suffices to know $\la \sigma n_2 \ra$ only which is given by 
\[
\langle\sigma n_2\rangle = \langle \sigma n_1\rangle\,\sinh \left(\left(\frac{L+1}{2}-2\right)k\right)/\sinh\left(\frac{L-1}{2}k \right)
\]
In the limit of thermodynamic system size $L \to \infty,$ this reduces to,
\bea
\langle\sigma n_2\rangle = e^{-k}\langle \sigma n_1\rangle \label{eq:s_n2}
\eea
with $k = \log (r + 1 + \sqrt{r^2 + 2r})$ taken positive. Combining \eref{e2}, \eref{eq:s_n2} and \eref{eq:n_ns},  we get two independent equations involving  $\rho$ and $\la \sigma n_1\ra,$
\bea
(2 e^{-k}-S-4r -2) \la \sigma n_1 \ra - D \rho &=& \gamma -\alpha \cr
D \la \sigma n_1 \ra + S\,\rho &=& \alpha + \gamma
\eea
which can be solved immediately to obtain,
\begin{equation}
\langle \sigma n_1 \rangle = \frac{\alpha\delta-\beta\gamma}{2(\alpha+\beta)(\gamma+\delta)+SR}
\end{equation}
and
\begin{equation}\label{den}
\rho  = \frac{(\alpha-\gamma)D - (S+4r+ 2(1-e^{-k})(\alpha+\gamma)}{D^2 - (S+4r+ 2(1-e^{-k}))S}
\end{equation}
The final formula \eref{exr} follows by using $2r+1-e^{-k} = R =r + \sqrt{r(r+2)}$.
%

\section{Density and current in rASEP for $r \uparrow \infty$}\label{rasep-infty}

For $r\rightarrow\infty$ and $E>0$, the results become independent of $\sigma$ and we have
 \begin{equation}
 \lambda_{\textrm{in}} = \frac{\alpha+\gamma}{2}, \, \, \lambda_{\textrm{out}} = \frac{\beta+\delta}{2} .
 \end{equation}
 In this limit, the model becomes a standard ASEP~\cite{sasamoto-asep}, but with the following boundary rates:
 
 \begin{itemize}
 \item entry at left: $\alpha_{\textrm{eff}}=\frac{\alpha+\gamma}{2}$,
 \;\;
 entry at right: $\gamma_{\textrm{eff}}=\frac{\alpha+\gamma}{2} e^{-E}$
 \item exit through left: $\beta_{\textrm{eff}}=\frac{\beta+\delta}{2} e^{-E}$,
 \;\; exit through right: $\delta_{\textrm{eff}}=\frac{\beta+\delta}{2}$
 \end{itemize}
 Defining
 \begin{equation}\label{asep-x1}
 x_1 := \frac{1}{2\alpha_{\textrm{eff}}} \left[1-e^{-E}-\alpha_{\textrm{eff}}+\beta_{\textrm{eff}}+\sqrt{(1-e^{-E}-\alpha_{\textrm{eff}}+\beta_{\textrm{eff}})^2+4\alpha_{\textrm{eff}}\beta_{\textrm{eff}}} \right]
 \end{equation}
 \begin{equation}\label{asep-x2}
 x_2 := \frac{1}{2\delta_{\textrm{eff}}} \left[1-e^{-E}-\delta_{\textrm{eff}}+\gamma_{\textrm{eff}}+\sqrt{(1-e^{-E}-\delta_{\textrm{eff}}+\gamma_{\textrm{eff}})^2+4\delta_{\textrm{eff}}\gamma_{\textrm{eff}}} \right]
 \end{equation}
we can use~\cite{sasamoto-asep} to get a simple expression for the average bulk current $j$ and the average bulk density $\rho$. There are three possible phases:
 \begin{itemize}
 \item $x_1>1, x_1>x_2$ : Low density phase with $\rho=\frac{1}{1+x_1}$ and $j \simeq (1-e^{-E})\frac{x_1}{(1+x_1)^2}$
 \item $x_2>1, x_2>x_1$ : High density phase with $\rho = \frac{x_2}{1+x_2}$ and $j\simeq(1-e^{-E})\frac{x_2}{(1+x_2)^2}$
 \item $x_1<1, x_2< 1$ : Maximal current phase with $\rho=0.5$ and $j \simeq \frac{1-e^{-E}}{4}$  
 \end{itemize}

\end{document}